\documentclass[twocolumn, amsmath, twocolappendix]{aastex631}
\usepackage{amsmath}
\usepackage{orcidlink}

\begin{document}

\title{Search for synchrotron pair echo emission following KM3-230213A}

\author{Angelina Sherman \orcidlink{0009-0007-2369-258X}}
\affiliation{Department of Physics, Wisconsin IceCube Particle Astrophysics Center, University of Wisconsin, Madison, WI, 53706}

\author{Nestor Mirabal \orcidlink{0000-0002-7021-5838}}
\affiliation{Center for Space Sciences and Technology, University of Maryland, Baltimore County, Baltimore, MD 21250}
\affiliation{Astrophysics Science Division, NASA Goddard Space Flight Center, Greenbelt, MD 20771, USA}
\affiliation{Center for Research and Exploration in Space Science and Technology, NASA Goddard Space Flight Center, Greenbelt, MD 20771}

\author{David Guevel \orcidlink{0000-0002-0870-2328}}
\affiliation{Department of Physics, Wisconsin IceCube Particle Astrophysics Center, University of Wisconsin, Madison, WI, 53706}

\author{Ke Fang \orcidlink{0000-0002-5387-8138}}
\affiliation{Department of Physics, Wisconsin IceCube Particle Astrophysics Center, University of Wisconsin, Madison, WI, 53706}

\author{Kohta Murase \orcidlink{0000-0002-5358-5642}}
\affiliation{Department of Physics; Department of Astronomy and Astrophysics; Center for Multimessenger Astrophysics, Institute for Gravitation and the Cosmos, The Pennsylvania State University, University Park, Pennsylvania 16802, USA}
\affiliation{Center for Gravitational Physics and Quantum Information, Yukawa Institute for Theoretical Physics, Kyoto, Kyoto 606-8502 Japan}

\author{Elizabeth Hays \orcidlink{0000-0002-8172-593X}}
\affiliation{Astrophysics Science Division, NASA Goddard Space Flight Center, Greenbelt, MD 20771, USA}

\begin{abstract}
The KM3NeT Collaboration has recently reported the detection of an extraordinary ultra-high-energy neutrino event with an energy of 220 PeV. Ultrahigh energy neutrinos and gamma-rays are co-produced in ultrahigh energy cosmic-ray interactions. If a UHE neutrino was produced within the large-scale structure around the source where it was accelerated, gamma-ray emission may be expected via the synchrotron pair echo mechanism. Here, we develop the synchrotron pair echo model in the specific context of the KM3NeT neutrino. Motivated by the fact that the synchrotron pair echo signal is expected to peak in the GeV - TeV band, and that the signal may appear as a dim, transient source,  
%that increases in brightness over time,  
we investigate the data collected by the Large Area Telescope (LAT) on-board the \textit{Fermi} Gamma-ray Space Telescope for transient and sub-threshold gamma-ray sources in the vicinity of the KM3NeT neutrino. We find three sub-threshold sources with TS $\gtrsim 16$ within $3.5^{\circ}$ of the neutrino event not included in any existing \textit{Fermi}-LAT catalogs, but note that none of the identified sub-threshold sources seem to be compelling candidates for synchrotron pair echo emission. 
\end{abstract} %%%%%%%%%

\section{Introduction}

The KM3NeT collaboration has recently reported the detection of an extraordinary ultra-high-energy event, KM3-230213A.  The event, which occurred on 13 February 2023, had properties of a muon with an estimated energy of $\sim$120 PeV. Because of its horizontal angle and very high energy, the muon likely originated from an astrophysical neutrino that interacted in the vicinity of the detector. The corresponding neutrino energy of $\sim$220 PeV makes KM3-230213A the highest-energy neutrino observed to date \citep{2025Natur.638..376K}. 

Ultrahigh-energy (UHE; $>100$~PeV) neutrinos and gamma-rays are the predicted counterparts of ultrahigh-energy cosmic rays, which are charged nuclei whose energies can exceed $10^{18}$~eV. While the sources that accelerate ultrahigh-energy cosmic rays remain unknown, observations of neutrinos and gamma-rays co-produced in ultrahigh-energy cosmic ray interactions may be used to better understand their origins. These secondary UHE neutrinos and gamma-rays may be produced inside a source (astrophysical origin) or by interactions of ultrahigh-energy cosmic rays during propagation through the cosmic microwave background (CMB) (cosmogenic origin; \citet{UHE_neutrinos_review}). Unfortunately, initial follow-up efforts have not succeeded in localizing a clear counterpart to KM3-230213A. \citet{2025arXiv250208484K} considered known blazars as candidate sources of KM3-230213A. They found four 4FGL-DR4 objects within the 99\% confidence region of the neutrino event, though none of them exhibit a gamma-ray flare coinciding with or after the neutrino arrival time. In addition, the High Altitude Water Cherenkov (HAWC) gamma-ray observatory conducted follow-up searches for TeV gamma-ray emission at the position of KM3-230213A on three time intervals after the neutrino observation, but observed no significant gamma-ray emission \citep{HAWC_ul}. 

UHE gamma-rays are difficult to detect because they are quickly attenuated by cosmic photon backgrounds. However, in the case of cosmogenic production, the UHE gamma-rays may generate an electromagnetic cascade (e.g., \citealp{1996ApJ...464L..75W,Murase:2009ah,2023MNRAS.519L..85M,2025ApJ...982L..16F}) resulting in an observable GeV-TeV signal. If KM3-230213A was cosmogenic, i.e. produced by a cosmic-ray interaction with the CMB during propagation, the photon co-produced with the neutrino would interact with the CMB to produce an electron-positron pair. The electron and positron would in turn inverse Compton scatter off photons in the CMB to produce a new generation of gamma-rays. This process would continue until the photon energy has dropped below the threshold for pair production off the CMB and the extragalactic background light (EBL), resulting in an electromagnetic cascade. This electromagnetic cascade would exhibit a distinct spectral shape that can separate the observation from a traditional gamma-ray transient such as a blazar flare. Deflections by electrons and positrons in the intergalactic magnetic field would result in an angular deflection and time delay for the observed cascade signature, and this signature is called an inverse-Compton pair echo~\citep[e.g.,][]{1995Natur.374..430P,1996ApJ...464L..75W,2002ApJ...580L...7D,Ichiki:2007nd,Murase:2008pe,Murase:2009ah}.  The inverse-Compton pair echo model for the case of KM3-230213A is developed in \citet{2025ApJ...982L..16F}. \citet{Milena_Cascades} searched for an inverse-Compton pair echo gamma-ray signature following the neutrino, but found no such emission. 

The inverse-Compton pair echo model assumes that the UHE neutrino was produced in the cosmic void where the magnetic field strength is low, far away from its accelerator source. On the other hand, if KM3-230213A was produced by a hadronic interaction within the large-scale structure around the source where its parent cosmic ray was accelerated (for example, a filament or the outskirt of a galaxy cluster), gamma-ray emission may be expected via the synchrotron pair echo mechanism. This mechanism is described in \citet{Murase:2011yw} and recently applied to the observation of gamma-ray bursts (e.g., \citealp{2025arXiv250415890D}). In this work, we develop the synchrotron pair echo mechanism for the case of KM3-230213A. 
%In both the synchrotron pair echo and inverse Compton cascade models, the gamma-ray signal is expected to peak in the GeV-TeV band, and may appear as a dim source. 
Motivated by the fact that the synchrotron pair echo signal may appear as a dim source in the GeV-TeV band, we investigate the \textit{Fermi}-LAT data in the vicinity of KM3-230213A for uncatalogued, sub-threshold\footnote{We consider sub-threshold sources to be those observed with less than $5\sigma$ significance and therefore not listed in the existing \textit{Fermi}-LAT catalogs.} sources. We find three sub-threshold sources of test statistic (TS) $\gtrsim 16$ within $3.5^\circ $ of the neutrino event (note that this is relaxed beyond the 99\% containment region of the neutrino, which is $3^\circ$). We define the TS in Section \ref{analysis}.

In Section \ref{synchrotron_model}, we develop the synchrotron pair echo model for the case of KM3-230213A. In Section \ref{analysis}, we describe our analysis procedure using the \textit{Fermi}-LAT data. In Section \ref{results}, we present details of the three sub-threshold gamma-ray sources uncovered by our analysis. %We also place limits on the magnetic field strength in the region where the neutrino was produced if it is assumed that none of these sources are a synchrotron pair echo signal. 

\section{Synchrotron Pair Echo Emission}\label{synchrotron_model}
If the UHE neutrino was produced within the large-scale structure around the source where it was accelerated, gamma-ray emission may be expected via the synchrotron pair echo mechanism. Assuming a neutrino energy of $\sim$220 PeV, the gamma-ray co-produced with the neutrino would have an energy of $E_\gamma \sim 2E_\nu \sim 440$ PeV. This gamma-ray would interact with the surrounding photon field and produce electrons through pair production; for the CMB alone, the interaction length for pair production would be $\lambda_{\gamma\gamma}\approx 170$~kpc \citep{2025ApJ...982L..16F}. If the magnetic field in the region is sufficiently strong, the resulting electron-positron pair remain confined to the magnetized structure and produce synchrotron radiation. Such a mechanism was proposed in \citet{Murase:2011yw}, and has been applied to the observations of blazars \citealp{Dermer:2012rg} and gamma-ray bursts (e.g., \citealp{2025arXiv250415890D}). We extrapolate this mechanism to KM3-230213A here. 

The secondary electrons, with energy $E_e \sim 220$~PeV ($\gamma_e = 4.3\times 10^{11}$), lose their energy to synchrotron radiation over the synchrotron cooling length scale, which is given by: 
\begin{equation}
    D_{\rm syn} = \frac{E_e}{(4/3) \sigma_T u_B \gamma_e^2},
\end{equation}
with $\sigma_T$ being the Thomson cross section and $u_B = B^2/8\pi$ being the magnetic energy density. Approximating the target photon field with a monoenergetic spectrum, the same electron would lose energy to inverse Compton scattering over the energy loss length: 
\begin{equation}
    D_{\text{IC}} = \frac{E_e}{(4/3) \sigma_T u_\text{rad} \gamma_e^2 f_{\text{KN}}},
\end{equation} 
where $u_\text{rad}$ is the energy density of the photon field in the electron's environment and $f_{\text{KN}} \approx (1+4\gamma_e\epsilon/m_ec^2)^{-1.5}$ for a mono-energetic target radiation field with energy $\epsilon$ \citep{Moderksi:2005jw}. 

\begin{figure}[t]
     \centering
    \includegraphics[width=0.49\textwidth]{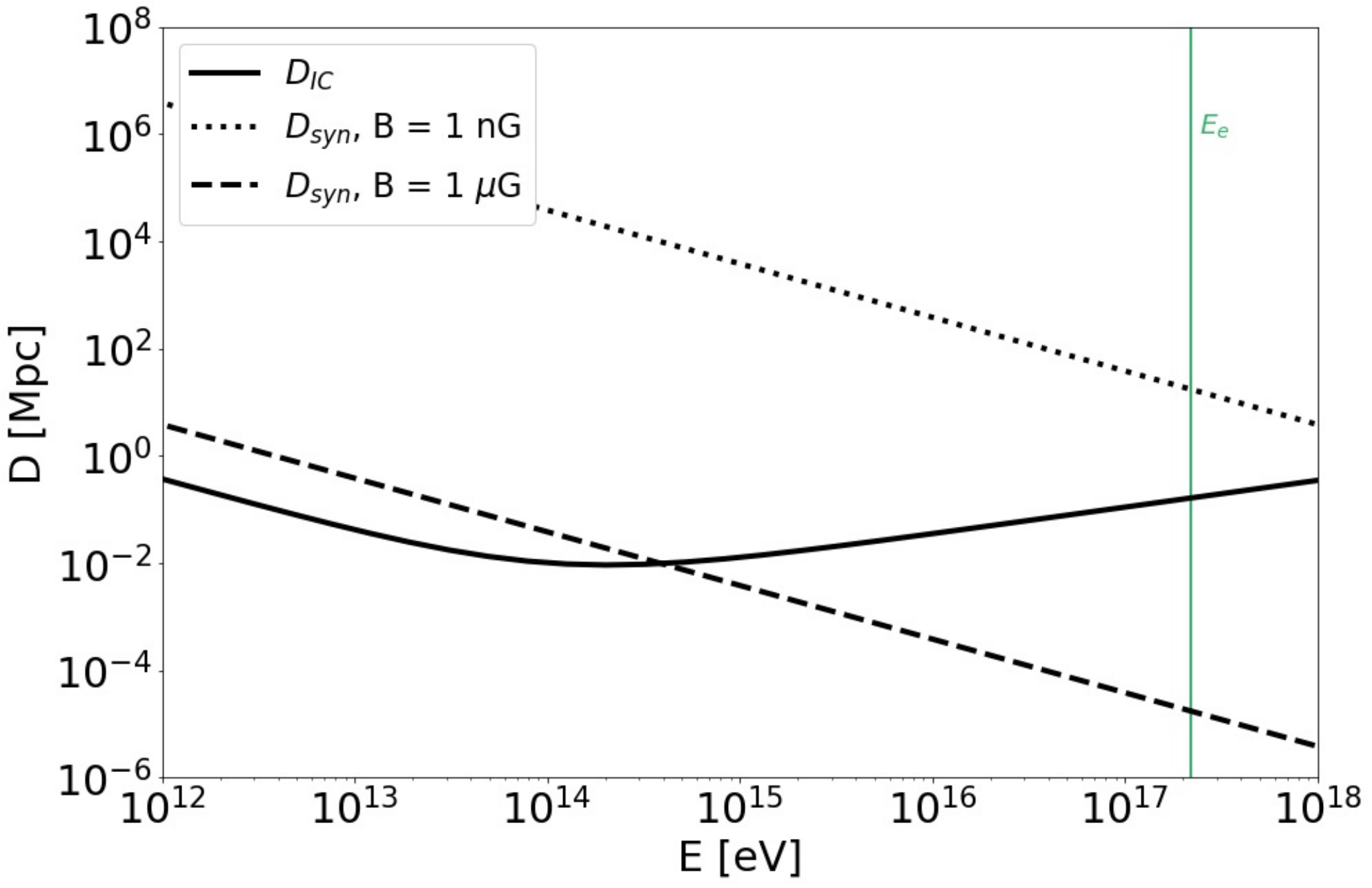}
    \caption{Comparison of inverse Compton (solid) and synchrotron (dashed/dotted) cooling length scales. The inverse Compton curve assumes a monoenergetic CMB spectrum; the synchrotron cooling lines correspond to different assumptions of the magnetic field strength. For the $\sim 220$ PeV electron-positron pair produced by the initial $440$ PeV gamma-ray interaction, the synchrotron echo mechanism may dominate over the inverse-Compton cascade for magnetic field strengths $\gtrsim 10$ nG. %{\ke [The synchrotron lines look weird. As $D_{\rm syn} \propto B^{-2}$, I'd expect a difference of 2.5e5 and 1e6 from 1~nG to  the   0.5 and 1~$\mu$G lines. ]} 
    }
    \label{cooling_lengths}
\end{figure}

A comparison of the synchrotron and inverse Compton cooling lengths is shown in Figure \ref{cooling_lengths}. For the case of the gamma-ray co-produced with KM3-230213A, and taking the target field to be the CMB alone ($u_\text{CMB} = 0.26~\text{eV}/\text{cm}^3$, $\epsilon_{\text{CMB}} = 6.3\times10^{-4}$ eV), the inverse Compton cooling length is $D_\text{IC, CMB} \sim 160$ kpc. Comparing this to the synchrotron cooling length indicates that the lowest magnetic field strength for which the regime may transition from inverse Compton cascade to synchrotron echo emission is $B \sim 10$ nG. If the inverse Compton energy loss length is shorter due to a denser photon field, a higher magnetic field strength would be required for synchrotron pair echo emission. In general, the average magnetic field strength in the large scale structure varies from $\sim$30 nG for filaments \citep{Carretti:2022tbj} and up to $\sim 40~\mu\text{G}$ for galaxy clusters \citep{Carilli_2002}. 

The synchrotron radiation emitted by the secondary electrons peaks at 
\begin{equation}
    E_{\rm syn} = \frac{3}{2}\frac{\hbar eB}{m_ec}\gamma_e^2 \langle \sin(\alpha_p)\rangle,
\end{equation}
where $\alpha_p$ is the pitch angle of the electron with respect to the magnetic field. For secondary electrons produced alongside KM3-230213A this ranges from $\sim 0.6$ GeV for a $0.2~\mu$G field and up to $\sim 100$ GeV for a 40 $\mu$G field, falling within the sensitivity of \textit{Fermi}-LAT.

Assuming that the field is coherent over the electron cooling length, the typical deflection angle of the electrons in the magnetic field is 
\begin{equation}
    \theta_e \approx %\frac{4}{\pi}
    \frac{D_{\rm syn}}{r_L},
\end{equation}
where $r_L = E_e/eB\langle \sin(\alpha_p)\rangle$ is the electron's Larmor radius.  
As a result of the electron's deflection, the synchrotron photons that reach the observer are separated from the source by an angle of  
\begin{equation}
    \label{syn_theta}
    \phi_\gamma \approx \frac{\lambda_{\gamma\gamma} + D_{\rm syn}}{d_s}\theta_e,
\end{equation}
where $\lambda_{\gamma\gamma}$ is the interaction length due to electron-positron pair production (see, e.g., Figure~1 of \citealp{2025ApJ...982L..16F}) and $d_s$ is the source distance. The observed angular deflection depends both on source distance and magnetic field in which the neutrino and gamma-ray were produced, but for the cases we consider will generally be pointlike: $\phi_\gamma \sim 0.2^\circ$ for source 10 Mpc away and 0.2 $\mu$G magnetic field to $\phi_\gamma \sim 0.01''$ for a source 5000 Mpc away and 40 $\mu$G magnetic field. Note that our simplified estimate for $\phi_\gamma$ depends on a small angle approximation, and does not apply if the emission is predicted to be very extended.  

The time delay over which the synchrotron photons arrive after the neutrino observation is~\citep{Murase:2011yw} 
\begin{equation}
    \Delta t \approx (\lambda_{\gamma\gamma} + D_{\text{syn}}) \frac{\theta_e^2}{2c}, %\simeq 53\,\,\rm yr, 
\end{equation}
where we have assumed that $\lambda_{\gamma\gamma}$ is smaller than the size of the large scale structure in which the interaction occurs. Taking $\lambda_{\gamma\gamma} = \lambda_{\gamma\gamma, \text{CMB}} \sim 170$ kpc $(E_\gamma/440~\rm PeV)$, the time delay ranges from 1 Myr for $\sim 30$ nG magnetic fields to 0.5 yr for $\sim 40~\mu$G magnetic fields. Note that $\lambda_{\gamma\gamma}$ may be much lower in dense infrared photon fields (for example, as low as $\sim1-10$~pc due to the infrared field of active galactic nuclei \citealp{Dermer:2012rg}); a lower value of $\lambda_{\gamma\gamma}$ would result in a smaller source extension and lower expected time spread for the synchrotron echo signal.

%In general, the observed synchrotron emission is likely to appear first as a faint signal and grow in intensity until it reaches a peak brightness at a time $\Delta t$ after the observation of the neutrino. 
The fraction of events arriving within some observation window  $t_\text{obs}$ is related to the total number of expected events by:
\begin{equation}
\label{distribution_function}
E^2\frac{dN}{dE }(t_\text{obs}) =  P(t_{\text{obs}},  ~\Delta t)\bigg[E_\gamma^2 \frac{dN}{dE_\gamma}\bigg]_{\text{total}}.
\end{equation}
The distribution function $P(t_{\text{obs}}, ~\Delta t)$ depends on the energy distribution of the primary photons as well as the strength and structure of the magnetic field, and may need to be evaluated numerically. Similar calculations for  echo emission mechanisms can be found in \citet{Ichiki:2007nd,Eskenasy:2022aup,Carpio:2022sml}.

\section{LAT Analysis} \label{analysis}

Motivated by the fact that the synchrotron echo emission corresponding to KM3-230213A is expected on the \textit{Fermi}-LAT sensitivity range, 
%and that the emission is likely to first appear as as faint source that grows in brightness over time, 
we perform a sub-threshold source search in the vicinity of KM3-230213A, in particular searching for emission that appears after the observation of the neutrino.
 
 We utilized the Fermitools\footnote{\url{https://fermi.gsfc.nasa.gov/ssc/data/analysis/documentation/}} package (v.2.2.0) and \texttt{fermipy} package (v.~1.3.0) \citep{fermipy} to search for potential synchrotron echo emission associated with KM3-230213A. We analyze events from 1 GeV to 1 TeV, and consider datasets over two different time intervals: the first interval is from MJD = 59988.00 to MJD = 60766.00, which corresponds to February 13, 2023 (the time of arrival of KM3-230213A) until April 1, 2025. The second dataset covers the time range from MJD = 54683.00 to MJD = 60766.00, which corresponds to all data taken by {\it Fermi}-LAT from August 5, 2008 until April 1, 2025. We selected events from the Pass 8 SOURCE class \citep{Pass8_Atwood, Pass8_Bruel} within a $15\times15$ degrees Radius of Interest (RoI) centered at the best-fit position of KM3-230213A and applied a maximum zenith angle cut of $90^{\circ}$ to minimize contamination from the Earth's limb. The \texttt{P8R3\_SOURCE\_V3} instrumental response functions were adopted, and additional contributions from nearby point sources were accounted for using the 4FGL-DR4 \citep{4FGL_DR4} catalog and standard background templates for the Galactic diffuse emission {\tt (gll\_iem\_v07.fits)}\footnote{\url{https://fermi.gsfc.nasa.gov/ssc/data/analysis/software/aux/4fgl/Galactic_Diffuse_Emission_Model_for_the_4FGL_Catalog_Analysis.pdf}} and the isotropic component {\tt (iso\_P8R3\_SOURCE\_V3\_v1.txt)}\footnote{The Galactic and isotropic diffuse emission templates can be downloaded at \url{https://fermi.gsfc.nasa.gov/ssc/data/access/lat/BackgroundModels.html}.} within $15^{\circ}$ of the RoI center. The RoI has pixel sizes of $0.1^{\circ}$.

In order to identify potential excess emission in the region of interest, we generated a test-statistic (TS) map using the \texttt{fermipy} function \texttt{gta.tsmap}. This function produces a TS map by introducing a tentative point source at every position in the region and calculating the likelihood ratio $\left(\frac{\mathcal{L}_0}{\mathcal{L}_s}\right)$ between the previously optimized model and the best-fit model with the additional test source. Here, $\mathcal{L}_0$ represents the likelihood of the original model and  $\mathcal{L}_s$  indicates  the likelihood of the model with the added source. The test statistic is then defined as TS \(= -2~{\rm log} \left(\frac{\mathcal{L}_0}{\mathcal{L}_s}\right)\).

To locate the most significant features in the region of interest, we first free the region using the \texttt{fermipy} function {\tt gta.free\_sources}, which frees the spectral parameters and normalization for all sources in the model. We then use the iterative source-finding algorithm {\tt gta.find\_sources} to identify emission peaks on the TS map. We constrain the positions of identified sub-threshold sources using {\tt gta.localize}, and use these coordinates to add each sub-threshold source to our model. To evaluate the test-statistic of a potential source using one degree of freedom, we allow the normalization of the isotropic and Galactic diffuse background components, as well as sources with TS $\geq 10$, to vary during fitting, but we hold the spectral parameters of all sources fixed (the candidate sub-threshold source's spectral index is fixed as $\Gamma = 2$, which is the default value in \texttt{fermipy}). Finally, we free all parameters in the region using \texttt{gta.free\_sources}  and generate a new fit with the added sub-threshold sources using {\tt gta.fit}. Using the output of this final fit, we generate  a light curve using \texttt{gta.lightcurve} and a TS Map using \texttt{gta.tsmap}. By default, \texttt{gta.tsmap} assumes that each test source is a point source with power law spectral index $\Gamma = 2$. The shape parameters of the test source and the parameters of background components remain fixed during this step of analysis (the background components are allowed to vary during the initial fit). For the spectral energy density (SED) analysis, we extend the energy range of our analysis to 500 MeV - 1 TeV. On this energy range, we free the region using \texttt{gta.free\_sources} and generate a \texttt{gta.fit} with the added sub-threshold sources. We use the output of this fit to evaluate the best-fit spectral index and expected energy flux of each sub-threshold source. Finally, we generate an SED for each sub-threshold source using  {\tt gta.sed}.\footnote{Example scripts for reproducing our analysis can be found at \url{https://github.com/angelinapartenheimer/Fermi-LAT-analysis-example}}

We relaxed our search criteria slightly beyond the 99\% error circle (radius $3^\circ$) defined by the KM3NeT collaboration, and instead consider all sources within $3.5^\circ $ of KM3-230213A; this ensures that any sources falling at the edge of the error circle or sources with extended emission are captured in the analysis. We select sub-threshold sources with TS $\gtrsim 16$ (which corresponds to roughly $3.5\sigma$ significance) that fall within this radius of KM3NeT. We perform separate searches using the two-year post-neutrino dataset and the all-time dataset. We identify three sub-threshold sources that meet these criteria, one of which is a transient sub-threshold gamma-ray source that appears only after the observation of KM3-230213A. 

\section{Results} \label{results}

\begin{table*}
    \begin{ruledtabular}
    \begin{tabular}{cccc}
    \vspace{0.1 pt} & \textbf{J0616.1-0428} & \textbf{J0614.6-0731} & \textbf{J0621.1-0610} \\
    \hline
    Dataset: & post-neutrino (2 year) & all-time (17-year) & all-time (17-year)\\
    \hline
    TS: & 19 & 16 & 21 \\
    \hline 
    Right ascension: & $94.02^\circ$ & $93.67^\circ$ & $95.28^\circ$ \\
    \hline
    Declination: & $-4.48^\circ$ & $-7.51^\circ$ & $-6.18^\circ$ \\
    \hline
    Localization (95\% conf): & $0.12^\circ$ & $0.18^\circ$ & $0.11^\circ$ \\
    \hline
    Offset: & $3.33^\circ$ & $0.69^\circ$ & $1.89^\circ$ \\
    \hline
    Spectral index: & $2.151 (\pm 0.02)$ & $2.068 (\pm 0.167)$ & $2.141 (\pm 0.142)$ \\
    \hline
    Energy flux (erg cm$^{-2}$ s$^{-1}$): & $3.75 \times 10^{-12} (\pm 2.2\times 10^{-13})$ & $1.05\times 10^{-12} (\pm 3.8\times 10^{-13})$ & $1.13\times 10^{-12} (\pm 3.4\times10^{-13})$ \\
    \hline
    Notable counterparts: & 1eRASS J061605.3-043303 & NVSS J061448-073001 & ICRF J062110.3-060954\\
    \vspace{0.1 pt} & NVSS J061605-043301 & 1eRASS J061439.9-073357\\
    \vspace{0.1 pt} & 2MASS J06160546-0432594\\
    \hline
    References: & \citet{microquasar_candidate} & \vspace{0.1 pt} & \citet{ICRF_radio} \\
    \vspace{0.1 pt} & \vspace{0.1 pt} & \vspace{0.1 pt} & \citet{Wibrals_2014}\\
    \hline
    Notes: & microquasar candidate & \vspace{0.1 pt} & radio blazar
    \end{tabular} 
    \end{ruledtabular}
    \caption{Detailed information for the three sub-threshold sources uncovered in our analysis. For each candidate source, the TS is evaluated assuming a fixed spectral index $\Gamma = 2$, while the spectral analysis and localization are performed after freeing all parameters in the region.}
    \label{sources}
\end{table*}

For each sub-threshold source, we searched for high-energy counterparts in a variety of source catalogs using the database source search tools such as Simbad \citep{Simbad}, the NASA/IPAC Extragalactic Database (NED\footnote{\url{https://doi.org/10.26132/NED1}}), 
and the XAMIN search tool provided by NASA's High Energy Astrophysics Science Archive Research Center. In particular, we considered sources in catalogs from the extended ROentgen Survey with an Imaging Telescope Array (eROSITA) \citep{eROSITA_DR1}, the NRAO 1.4 GHz VLA Sky Survey (NVSS) \citep{NVSS}, and the Two Micron All Sky Survey (2MASS) \citep{2MASS}, the Wide-field Infrared Survey Explorer (WISE) \citep{WISE}. While we describe the most notable counterparts that we find, these observations should be considered as purely informative and not as a formal counterpart search.

\begin{figure*}
     \centering
    \includegraphics[width=0.49\textwidth]{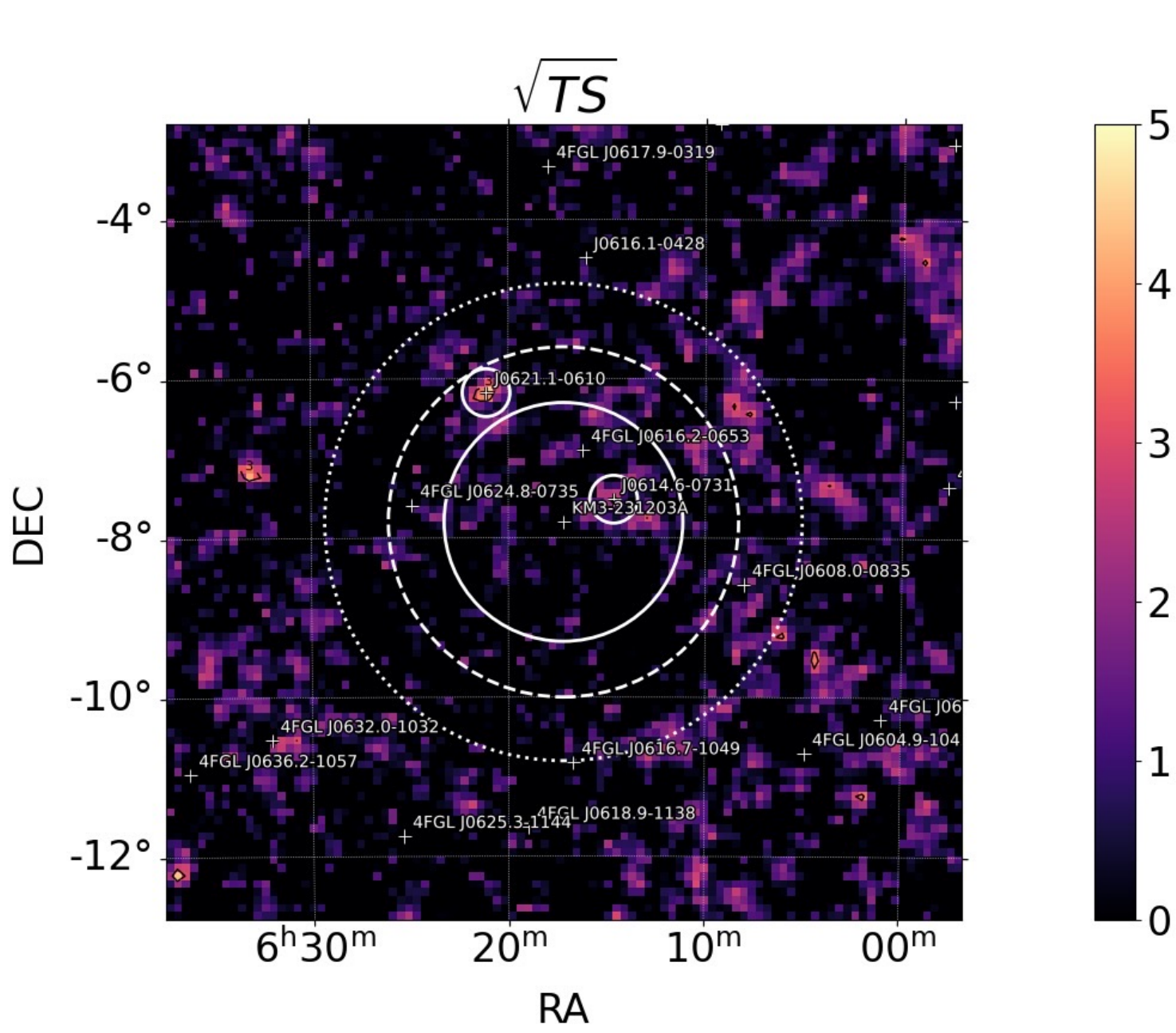}
    \includegraphics[width=0.49\textwidth]{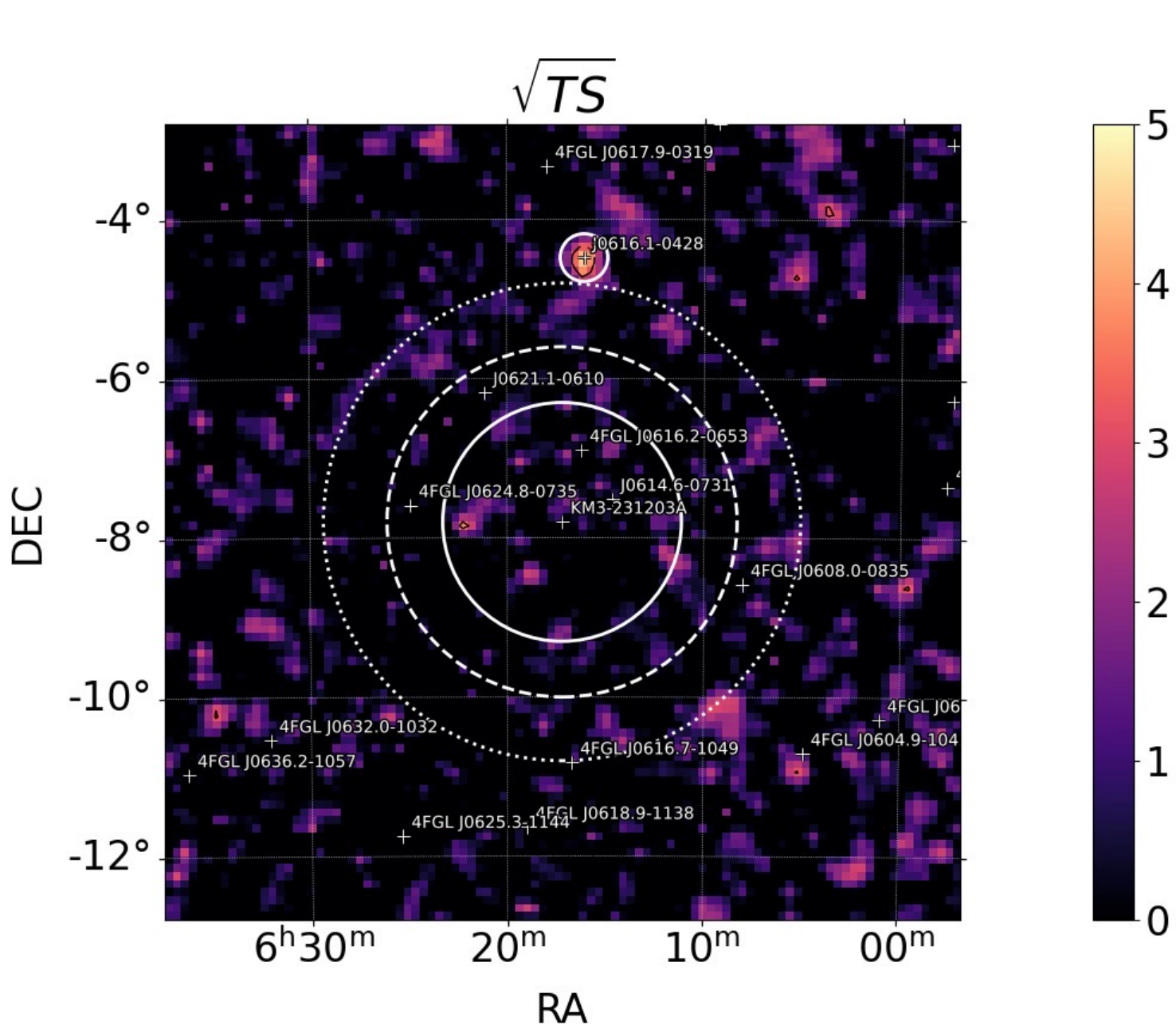}
    \caption{
    Left panel: \textit{Fermi}-LAT TS  map (1 GeV -- 1 TeV) of the $10^{\circ} \times 10^{\circ}$ region centered on KM3-230213 using 17  years of {\it Fermi}~LAT data prior to the neutrino detection showing the sub-threshold sources J0614.6-0731 and J0621.1-0610. Right panel: TS  map generated using LAT data from the time of the neutrino detection to slightly more than 2 years after the event showing the transient sub-threshold gamma-ray source J0616.1--0428, which appears only after the observation of KM3-230213A. The large concentric circles indicate the 68\%, 90\%, and 99\% error circles for the localization of KM3-230213A. In both the TS Maps, the circled sources are excluded from the model to highlight their emission. Both TS maps are fit assuming a point source model with a spectral index $\Gamma = 2$.}
    \label{tsmaps}
\end{figure*}

\begin{figure}[t]
     \centering
    \includegraphics[width=0.45\textwidth]{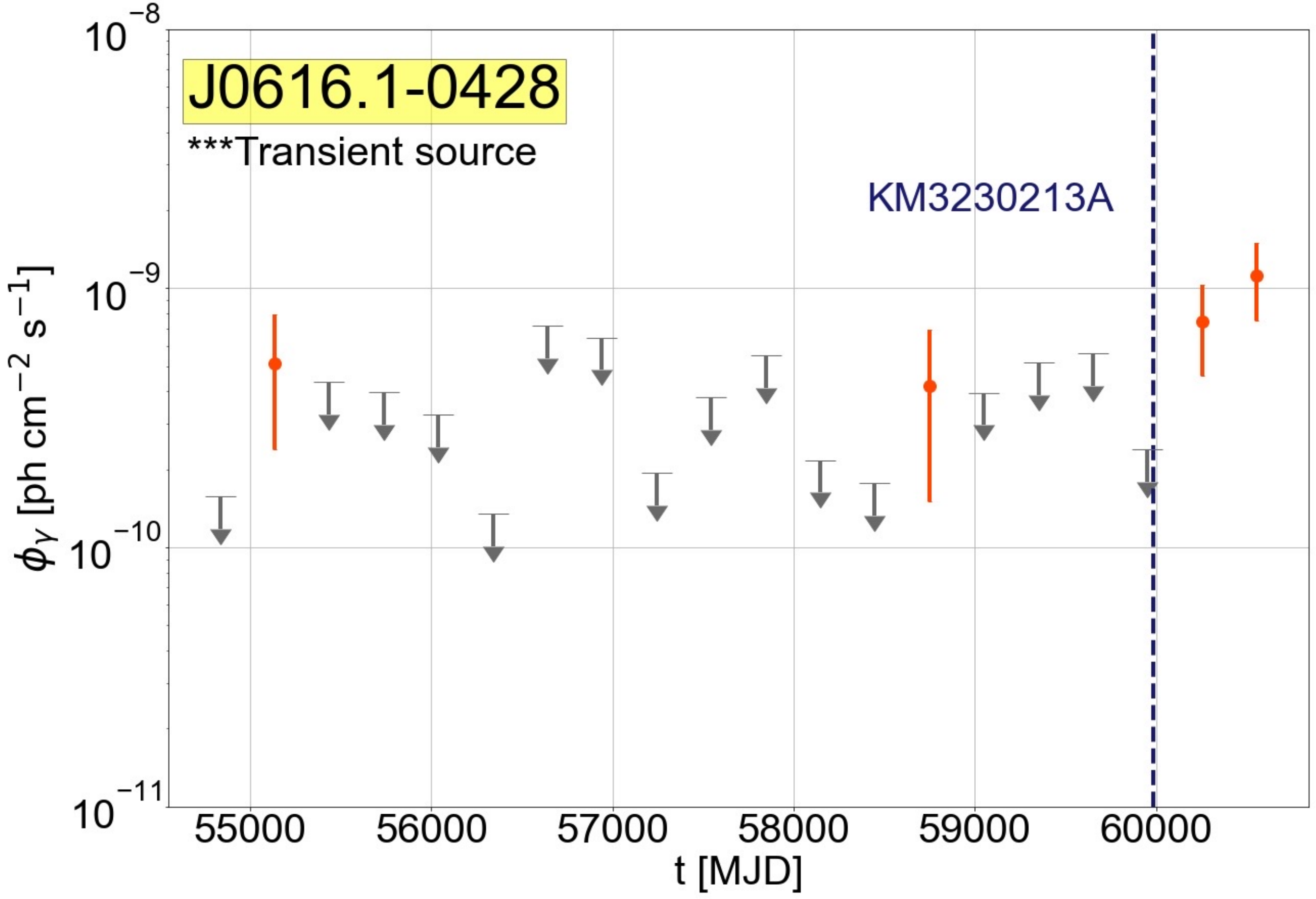}
    \includegraphics[width=0.45\textwidth]{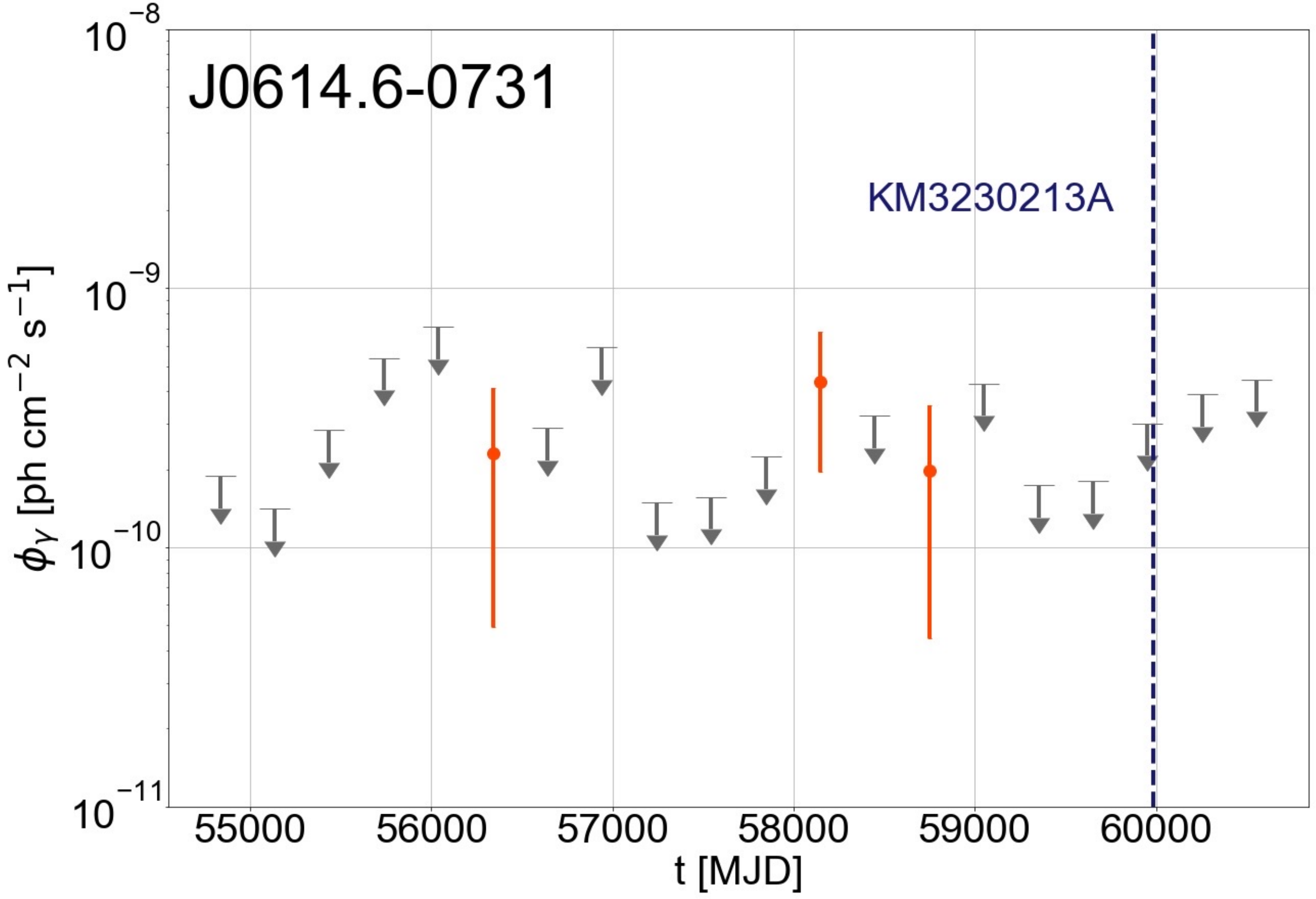}
    \includegraphics[width=0.45\textwidth]{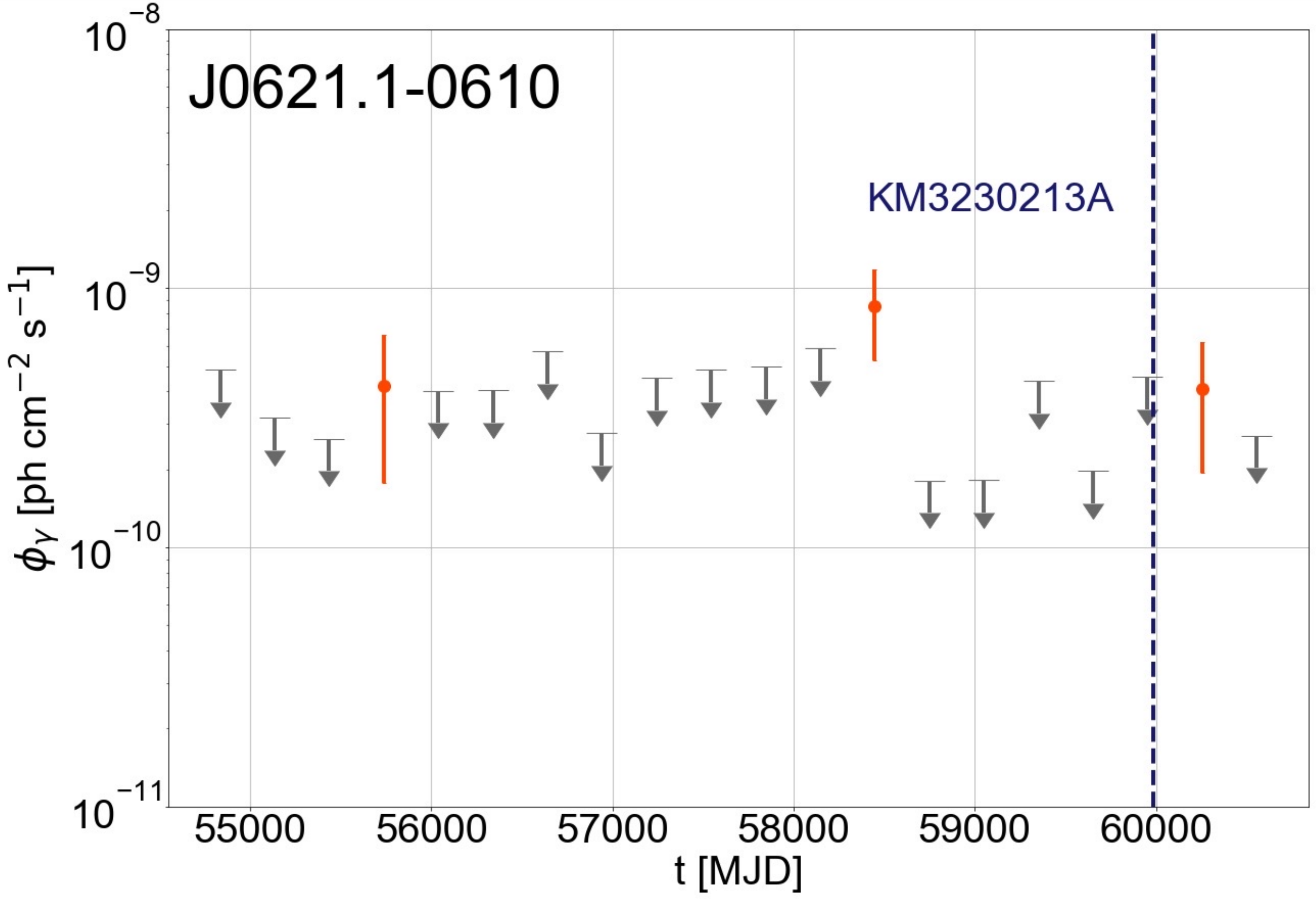}
    \caption{Light curves for each of the sub-threshold sources discussed in this work. The light curves are produced in the 1 GeV - 1 TeV energy range. Within each time bin, an errorbar is placed if TS $\geq 4$, while a $2\sigma$ upper limit is estimated when TS $> 4$. Each time bin is $\sim 300$ days wide. The vertical dashed line indicates the observation time of the KM3NeT neutrino KM3-230213A. Both J0616.1-0428 and J0621.1-0610 appear to fluctuate immediately after the observation of KM3-230213A, while J0614.6-0731 only has upward fluctuations years before the neutrino observation. Note that for a sub-threshold source, fluctuations above $2\sigma$ are not inconsistent with statistical variation and should not be considered as an accurate proxy of the source's activity.}
    \label{lightcurves}
\end{figure}

\subsection{Post-neutrino ($\sim$ 2) year dataset: a sub-threshold transient source}

\textbf{J0616.1-0428}: We identify one sub-threshold transient source that appears after the observation of KM3-230213A. This sub-threshold source is labeled J0616.1-0428 and has a TS $\sim 19$ (see Figure \ref{tsmaps}). Using {\tt gta.localize}, we find a position for the candidate source (J2000) RA: 06:16:05.52  Dec:--04:28:46.08 with  a 95\% confidence radius of 0.12 degrees, which corresponds to an offset of $\sim 3.3^\circ$ from KM3-230213A. The light curve is shown in Figure \ref{lightcurves} and reveals marginal initial activity approximately 200 days after the KM3NeT detection, with a more prominent peak occurring around 650 days post-neutrino event.

 The {\it Fermi}-LAT flux of this source on the time interval following the observation of KM3-230213A (from February 13, 2023 until April 1, 2025) is calculated to be  $3.75 \times 10^{-12} (\pm 2.2\times 10^{-13})$ erg cm$^{-2}$ s$^{-1}$ with a spectral index of $\Gamma=2.151 \pm 0.02$. The majority of the signal is detected at energies where the LAT effective area peaks, specifically between 1-10 GeV (see Figure \ref{spectra_alltime}). %While the emission is transient, it does not exhibit the characteristics of electromagnetic cascade radiation, as the observed $E^{-2.151}$ spectrum is much softer than what is expected for such a scenario.  
 Using the full $\sim 17$ year dataset, a point source at the position of J0616.1-0428 has a TS = 6, which indicates that the candidate source is a truly transient event and does not appear as a steady-state sub-threshold source during the full duration of \textit{Fermi}-LAT observation. The SED analysis using 17 years of LAT data prior to the neutrino event suggests an upper limit for the flux of $\leq 6.2  \times 10^{-13}$ erg cm$^{-2}$ s$^{-1}$ at the sub-threshold source's location.

The most interesting counterpart we find for this transient sub-threshold gamma-ray source is the eROSITA X-ray source 1eRASS J061605.3-043303, which is located 4.31' away from J0616.1-0428 and has a flux of $2.05\times10^{-13}$ erg/cm$^2$/s on the 0.2-2.3 keV band. The eROSITA source falls within less than 0.1' from the infrared source 2MASS J06160546-0432594 (magnitude $16.44\pm0.07$) and the radio point source NVSS J061605-043301 (which has a flux density of $6.9 \pm 0.5$ mJy on the 1.4 GHz band and is relatively faint). The radio point source also appears in the VLASS survey \citep{VLASS} as VLASS1QLCIR J061605.50-043259.2 with an integrated flux of $4.66\pm0.26$ mJy. Notably, the infrared source 2MASS J06160546-0432594 has been identified as a possible microquasar candidate in \citet{microquasar_candidate}\footnote{https://images.astronet.ru/pubd/2008/09/28/0001230826/447-452.pdf}. However, no spectroscopic confirmation has been reported in the literature. The counterpart is identified in the Gaia catalog as having a very high ($\sim 1$) probability of being a star, and a near-zero probability of being a quasar or galaxy. Microquasars have been known to flare in the GeV band \citep{Prokhorov_2022, Bodaghee_2013, 2025ApJ...979L..40M}, and if the candidate source is indeed a microquasar flaring in gamma-rays it would lack the sufficient bolometric luminosity to generate a neutrino with the energy of KM3-230213A. 

\subsection{Full ($\sim$ 17) year dataset: sub-threshold sources}

\textbf{J0614.6-0731}: The nearest sub-threshold source with respect to KM3-230213A is J0614.6-0731, which has a TS $\sim 16$. We localize the candidate source to a position of RA: 06:14:39.5  Dec:--07:30:51.7 with a 95\% confidence radius of $0.18^\circ$. This corresponds to an offset of $\sim 0.7^\circ$, which is within the 68\% error circle of KM3-230213A. This sub-threshold source falls 2.4' from the radio source NVSS J061448-073001 and 3.1' from the eROSITA X-ray source 1eRASS J061439.9-073357. The light curve indicates that any gamma-ray fluctuations occurred before KM3-230213A, with the most recent fluctuation preceding the neutrino by about three years. Performing a spectral analysis yields a spectral index of $\Gamma=2.068 \pm 0.167$ and an energy flux of $1.05\times 10^{-12} (\pm3.8\times10^{-13})$ erg cm$^{-2}$ s$^{-1}$. 

\textbf{J0621.1-0610}: The most significant sub-threshold source that appears in the vicinity of KM3-230213A is J0621.1-0610 with a TS $\sim 21$. We localize this candidate source to the position RA: 6:21:6.70  Dec:--06:10:33.3 with a 95\% confidence radius of $0.11^\circ$. This position corresponds to an offset of $\sim 1.9^\circ$ and is located within the 90\% error circle of KM3-230213A. The sub-threshold source appears to have had four upward fluctuations in gamma-rays in the past fifteen years, with the most recent fluctuation occurring coincident with the KM3-230213A (see Figure \ref{lightcurves}). Using only the $\sim 2$ year time interval following KM3-230213A, this candidate source has a TS of $\sim 13$, which falls slightly below our defined threshold of TS $\sim16$ and is the reason that we did not identify this as a transient sub-threshold source using the two-year dataset. The sub-threshold source has a spectral index of $\Gamma=2.141 \pm 0.142$ and a predicted energy flux of $1.13\times 10^{-12} (\pm 3.4 \times 10^{-13})$ erg cm$^{-2}$ s$^{-1}$. The emission peaks on 10 - 100 GeV (see Figure \ref{spectra_alltime}). 

J0621.1-0610 falls within 1' of ICRF J062110.3-060954, which is a radio-loud blazar \citep{ICRF_radio}. This source has an estimated integrated flux density of 66 mJy on the X-band and  81 mJy on the S-band. This source appears in other catalogs as WISE J062110.35-060954.0, 2MASS J06211034-0609539, and NVSS J062110-060954. It also appears in the WISE Blazar-like Radio-Loud Source catalog \citep{Wibrals_2014, Wibrals_2019} as WIBRaLS J0621-0609, where it is idenitified as a BL Lac.  

\smallskip
Note that in the light curves shown in Figure \ref{lightcurves}, many of the points indicating above $2\sigma$ observation are very close to the level of the upper limits and are consistent with background fluctuations. For the 20 bins comprising the light curve, a background observation is expected to have one bin fluctuate above the 95\% confidence level. For a source that is near threshold, upward fluctuations of three or four bins are consistent with statistical variation, and should not be considered good indicators of the source's activity.

\subsection{Constraint from nondetection}

 If we assume that no synchrotron echo emission was discovered in our analysis, we can place a limit on the time delay corresponding to the synchrotron echo signal following KM3-230213A. Because the time delay depends directly on the magnetic field strength in the environment where the neutrino was produced, a lower limit on the time delay may be used to place a constraint on the corresponding magnetic field strength. 
 
The expected gamma-ray fluence can be estimated directly from the neutrino observation. Based on the observation of KM3-230213A alone, the KM3NeT Collaboration inferred a full-sky averaged per-flavor isotropic neutrino flux of $E_\nu^2 \Phi_\nu = 5.8\times 10^{-8}\,\rm GeV \,cm^{-2}\,s^{-1}\,sr^{-1}$ \citep{2025Natur.638..376K}. This can be used to infer a neutrino fluence of $E_\nu^2 \frac{dN}{dE_\nu} \approx E_\nu^2 \Phi_\nu 4\pi \Delta T = 21~\text{GeV}/\text{cm}^2$, where $\Delta T = 335$ days is the livetime of ARCA used to derive the isotropic flux. We use this to estimate an expected gamma-ray fluence of $E_\gamma^2 \frac{dN}{dE_\gamma} \approx \frac{4}{3} E_\nu^2 \frac{dN}{dE_\nu}$ based on the photopion production scenario. 

We use the results of our analysis to infer a limit on the observed photon flux in the vicinity of the neutrino. Using the two-year gamma-ray dataset and assuming a source power law spectrum with spectral index $\Gamma=2$, for a test source at the position of KM3-230213A we evaluate a 95\% confidence level flux upper limit of $E_\gamma^2\frac{dN_\gamma}{dEdt} = 2.29\times10^{-12}\,\rm erg\,cm^{-2}\,s^{-1}$ on 1 GeV--1 TeV. This upper limit is evaluated during the final fitting stage of our analysis using \texttt{gta.fit}. 

\begin{figure}[t]
     \centering
    \includegraphics[width=0.45\textwidth]{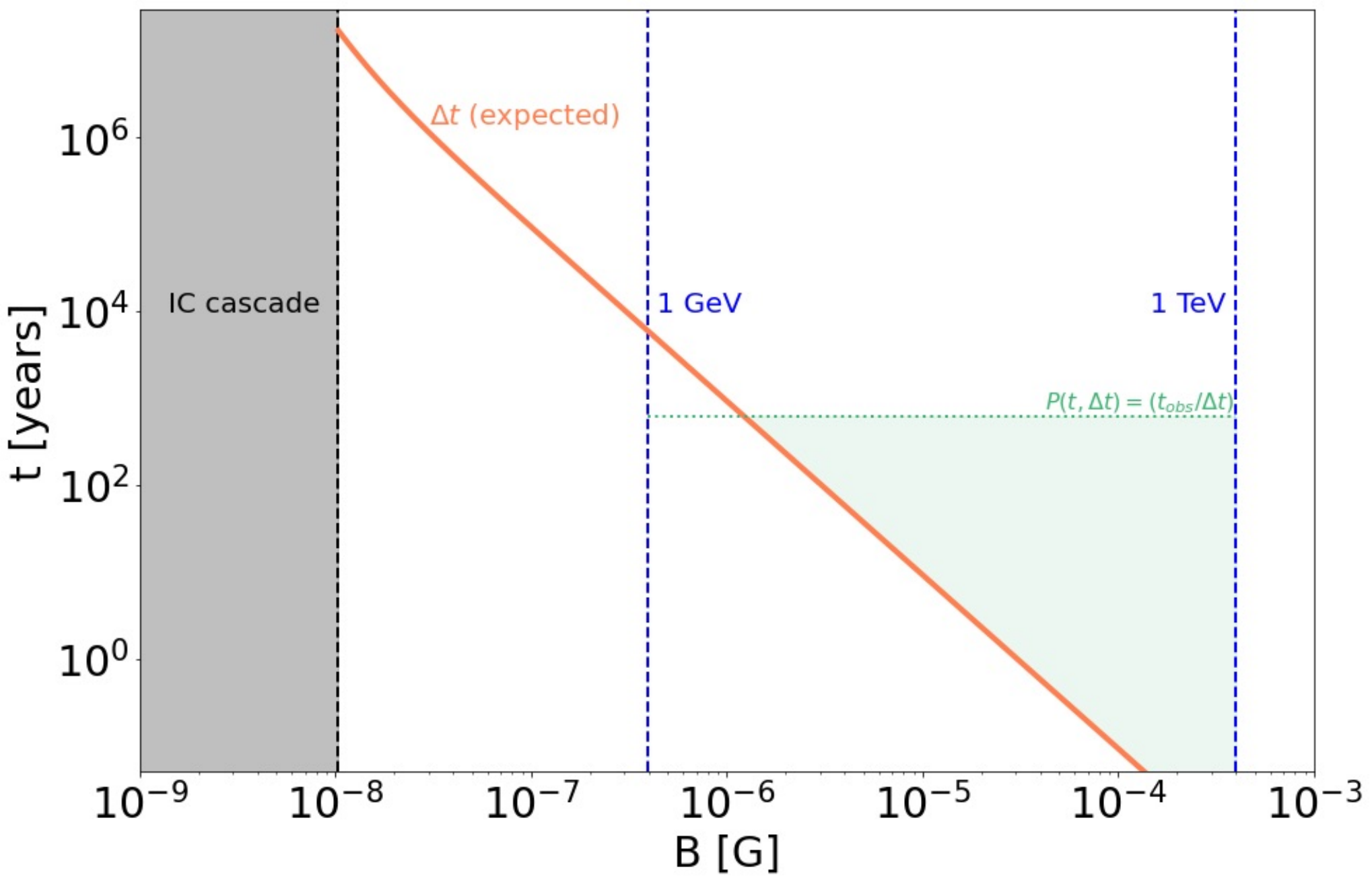}
    \caption{Lower limit (green) on $\Delta t$ found under the assumption that none of the sub-threshold sources uncovered in this work constitute synchrotron pair echo emission. Here, we make the simple assumption of a uniform increase in the expected photon flux. The orange line indicates the expected time delay for the synchrotron echo signal. %Note that we show a conservative estimate for the expected time delay as we consider pair production using the CMB only; using a denser photon field could yield to more efficient $\gamma\gamma$ pair production, which would correspond to a shorter time delay.
    }
    \label{constraint}
\end{figure}

Here, we assume a uniform accumulation of events: $P(t_{\text{obs}}, ~\Delta t) = \frac{t_{\text{obs}}}{\Delta t}$, corresponding to a constant photon flux \citep{Takahashi:2008pc, Murase:2008pe}. This gives a simple, order-of-magnitude estimate of the expected signal. We assume that the measured flux on the observation time interval $t_{\text{obs}}$ is $E_\gamma^2\frac{dN_\gamma}{dEdt} = \frac{1}{t_\text{obs}}E_\gamma^2\frac{dN_\gamma}{dE}(t_\text{obs})$. Inserting our measured upper limit on the gamma-ray flux and predicted total number of observed events into Equation \ref{distribution_function}, we infer a lower limit on the synchrotron echo time delay $\Delta t$ for our assumption of the distribution function. Our results are shown in Figure \ref{constraint}. Comparing our lower limit to the predicted synchrotron echo time delay shows that our analysis disfavors a scenario where the neutrino was produced in a section of large scale structure with magnetic field $\gtrsim 1~\mu$G.

\section{Summary and Discussion}

With an energy of 220 PeV, the ultrahigh energy neutrino event KM3-230213A would be co-produced with a gamma-ray photon of energy $\sim 440$ PeV. If the neutrino was produced far outside the large scale structure in the presence of low IGMF, a gamma-ray signal may be expected via the inverse-Compton pair echo mechanism. On the other hand, if the neutrino was produced near the source in the large scale structure where there is a stronger magnetic field, a gamma-ray signal may be expected by the synchrotron pair echo mechanism. In this work, we develop a model for synchrotron pair echo emission applied to the specific case of KM3-230213A. The synchrotron echo emission corresponding to the neutrino event is expected to appear on the GeV-TeV energy range, and may 
%lag the observation time of KM3-230213A by a duration of 
be spread over an interval of time on the order of months to years following the observation of KM3-230213A depending on the strength of the magnetic field in the environment where the gamma-ray interacts. 
%In general, such emission would likely appear as a dim source that becomes brighter with time. 
Motivated by this, we investigate the \textit{Fermi}-LAT data for sub-threshold emission in the vicinity of KM3-230213A, considering in particular emission that appears after the neutrino observation.

We used the \textit{Fermi}-LAT data to search for sub-threshold sources with TS $\gtrsim 16$ that fall within $3.5^\circ $ degrees of the KM3-230213A localization, a search area slightly relaxed beyond the 99\% containment region of $3^\circ$ for KM3-230213A. We searched two datasets separately: the first one uses the complete $\sim$ 17-year dataset from all \textit{Fermi}-LAT observations. The second one uses the $\sim 2$ years of data from the observation time of KM3-230213A until April 1, 2025. We uncovered 3 sub-threshold sources that are not catalogued by \textit{Fermi}-LAT. Two sub-threshold sources are detected on the all-time dataset, while one candidate source seems to be a transient sub-threshold gamma-ray source that becomes significant only after the observation of KM3-230213A. We note that for all the sub-threshold sources discussed in this work, the $> 2\sigma$ fluctuations observed in the light curves are consistent with natural statistical variation and are not a good indicator of the source's intrinsic variability.

The transient sub-threshold gamma-ray source (J0616.1--0428) is located just outside the 99\% error circle of KM3---230213A and is of interest because of its temporal coincidence with the observation of the neutrino. However, this sub-threshold source may be the unrelated flaring of a microquasar, in which case it would not have sufficient bolometric luminosity to produce a neutrino at hundreds of PeV. We note, however, that Fermi-LAT evidence for GeV emission in microquasars is limited to the three of the brightest microquasars ever observed \citep{2025ApJ...979L..40M}. While a firm identification of a microquasar as a gamma-ray emitter would be extremely interesting, it requires confirmation using correlated variability with a different wavelength. 

Another sub-threshold source that has an upward fluctuation after the neutrino observation is J0621.1-0610, which appears on the full $\sim$ 17-year dataset and falls within the 90\% error circle of KM3-230213A. While this candidate source has had three upward fluctuations in gamma-rays in the years preceding KM3-230213A, the most recent fluctuation occurs immediately after the neutrino observation. This sub-threshold source appears to be coincident with a known radio blazar, and is identified in \citet{Wibrals_2014} as a BL Lac. In general, \citet{Das_2025} notes that to reproduce the neutrino flux observed by KM3NeT, the required proton luminosity for sources with redshift $z \gtrsim 1$ exceeds the Eddington luminosity of typical AGN, implying that a cosmogenic origin from line-of-sight blazar emission is difficult to motivate unless the source is very powerful and close by. 

As a simple check to determine whether or not the number of sub-threshold sources that we identified around KM3-230213A was unusual, we performed a source search analysis with a $10^\circ \times 10^\circ$ patch of sky centered on RA: 19:36:48.00  Dec: -3:41:2.40, which is at the same Galactic latitude as KM3-230213A but $180^{\circ}$ away in Galactic longitude. Here we also find two sub-threshold sources with TS $< 16$ and offset less than $3.5^\circ$ from the RoI center, which indicates that the number of sub-threshold sources we find is roughly consistent with the number of sub-threshold sources expected in other sky regions of the same Galactic latitude. This test demonstrates that the number of sub-threshold sources uncovered around KM3-230213A could be found in a different sky region at the same latitude; in other words, the sub-threshold emission around KM3-230213A does not appear to be unusual or unique.

While we do not perform a statistically-robust assessment  of the merit of the identified sub-threshold sources as synchrotron echo emission, we note qualitatively that none of the sources appear to be promising candidates for this type of emission. Furthermore, we have noted that all `flaring' activity from the identified sub-threshold sources is consistent with background fluctuations, and that the number of sub-threshold sources identified in the region is consistent with another sky region at the same Galactic latitude. Assuming that we have not observed synchrotron pair echo emission, our results disfavor a scenario where the neutrino was produced in a section of large scale structure with larger than $\sim \mu$G magnetic fields. Such a magnetic field is higher than that typical for cosmic filaments, but is on the order of magnetic fields inferred for galaxy clusters. We note that our estimate for the expected time delay is conservative, as we consider pair production using the CMB only; using a denser photon field could yield to more efficient $\gamma\gamma$ pair production, which would correspond to a shorter time delay. On the other hand, our assumption of a uniform photon flux from the synchrotron pair echo is extremely simple, and choosing a different distribution function may result in a weaker constraint.

Our inferred limits apply only to a scenario where the neutrino was produced in large-scale-structure outside the source that accelerated its parent cosmic ray. 
If KM3-230213A was produced inside an astrophysical source, the corresponding gamma-ray signal may be attenuated if the source is radio loud. As is pointed out in \citet{2025ApJ...982L..16F}, for a photon at $\sim 440$ PeV, the pair production cross section would peak for interactions with radio photons at $\sim 140$ MHz, which would be detectable by low-frequency radio observatories. In the case that the neutrino was produced inside a source with a sufficiently luminous radio field around this frequency, the corresponding gamma-ray signal could be fully attenuated and may not be detected. Because of this, if no gamma-ray signature is observed to follow KM3-230213A, radio-loud sources, in particular those known to emit high-energy radiation, may be particularly interesting candidates for the neutrino's origin.

The observation of the ultrahigh-energy neutrino event KM3-230213A was unprecedented, and the lack of an easily-identified counterpart highlights the importance of detailed, multiwavelength follow-up searches. While this analysis only considered $\sim2$ years of observation following KM3-230213A, a delayed gamma-ray signature could still be anticipated in the coming years, strongly motivating continued monitoring in the GeV-TeV band. As a number of new observatories designed to target ultrahigh energy neutrinos are expected to begin operating in the next decade, an understanding of the secondary gamma-ray emission that may accompany ultrahigh-energy neutrinos is crucial to identifying the origins of these extremely energetic and mysterious particles.

\section*{Acknowledgments}
K.F. acknowledges support from the National Science Foundation (PHY-2238916) and the Sloan Research Fellowship. This work was supported by a grant from the Simons Foundation (00001470, KF).  We also acknowledge NSF Grants Nos.~AST-2108466 (K.M.), AST-2108467 (K.M.), AST-2308021 (K.M.), and KAKENHI No.~20H05852 (K.M.). 
The material is based upon work supported by NASA under award number 80GSFC24M0006. This research has made use of data and web tools obtained from the High Energy Astrophysics Science Archive Research Center (HEASARC), a service of the Astrophysics Science Division at NASA/GSFC and of the Smithsonian Astrophysical Observatory's High Energy Astrophysics Division. This research has also made use of the NASA/IPAC Extragalactic Database, which is funded by the National Aeronautics and Space Administration and operated by the California Institute of Technology.

The \textit{Fermi}-LAT Collaboration acknowledges generous ongoing support from a number of agencies and institutes that have supported both the development and the operation of the LAT as well as scientific data analysis. These include the National Aeronautics and Space Administration and the Department of Energy in the United States,the Commissariat à l'Energie Atomique and the Centre National de la Recherche Scientifique / Institut National de Physique Nucléaire et de Physique des Particules in France, the Agenzia Spaziale Italiana and the Istituto Nazionale di Fisica Nucleare in Italy, the Ministry of Education, Culture, Sports, Science and Technology (MEXT), High Energy Accelerator Research Organization (KEK) and Japan Aerospace Exploration Agency (JAXA) in Japan, and the K. A. Wallenberg Foundation, the Swedish Research Council and the Swedish National Space Board in Sweden. Additional support for science analysis during the operations phase from the following agencies is also gratefully acknowledged: the Istituto Nazionale di Astrofisica in Italy and the Centre Nationald'Etudes Spatiales in France. This work performed in part under DOE Contract DE-AC02-76SF00515.

 %\bibliography{reference}

\begin{thebibliography}{}
\expandafter\ifx\csname natexlab\endcsname\relax\def\natexlab#1{#1}\fi
\providecommand{\url}[1]{\href{#1}{#1}}
\providecommand{\dodoi}[1]{doi:~\href{http://doi.org/#1}{\nolinkurl{#1}}}
\providecommand{\doeprint}[1]{\href{http://ascl.net/#1}{\nolinkurl{http://ascl.net/#1}}}
\providecommand{\doarXiv}[1]{\href{https://arxiv.org/abs/#1}{\nolinkurl{https://arxiv.org/abs/#1}}}

\bibitem[{Ackermann {et~al.}(2022)Ackermann, Agarwalla, Alvarez-Muñiz,
  Batista, Argüelles, Bustamante, Clark, Cummings, Das, Decoene, Denton,
  Dornic, Dzhilkibaev, Farzan, Garcia, Garzelli, Glaser, Heijboer, Hörandel,
  Illuminati, Jeong, Kelley, Kelly, Kheirandish, Klein, Krizmanic, Larson, Lu,
  Murase, Narang, Otte, Prechelt, Prohira, Reno, Resconi, Santander, Suvorova,
  Valera, Vandenbroucke, Wiencke, Wissel, Yoshida, Yuan, Zas, Zhelnin, \&
  Zhou}]{UHE_neutrinos_review}
Ackermann, M., Agarwalla, S.~K., Alvarez-Muñiz, J., {et~al.} 2022, High-Energy
  and Ultra-High-Energy Neutrinos.
\newblock \doarXiv{2203.08096}

\bibitem[{{Atwood} {et~al.}(2013){Atwood}, {Albert}, {Baldini}, {Tinivella},
  {Bregeon}, {Pesce-Rollins}, {Sgr{\`o}}, {Bruel}, {Charles}, {Drlica-Wagner},
  {Franckowiak}, {Jogler}, {Rochester}, {Usher}, {Wood}, {Cohen-Tanugi}, \&
  {Zimmer}}]{Pass8_Atwood}
{Atwood}, W., {Albert}, A., {Baldini}, L., {et~al.} 2013, arXiv e-prints,
  arXiv:1303.3514, \dodoi{10.48550/arXiv.1303.3514}

\bibitem[{{Ballet} {et~al.}(2023){Ballet}, {Bruel}, {Burnett}, {Lott}, \& {The
  Fermi-LAT collaboration}}]{4FGL_DR4}
{Ballet}, J., {Bruel}, P., {Burnett}, T.~H., {Lott}, B., \& {The Fermi-LAT
  collaboration}. 2023, arXiv e-prints, arXiv:2307.12546,
  \dodoi{10.48550/arXiv.2307.12546}

\bibitem[{Bodaghee {et~al.}(2013)Bodaghee, Tomsick, Pottschmidt, Rodriguez,
  Wilms, \& Pooley}]{Bodaghee_2013}
Bodaghee, A., Tomsick, J.~A., Pottschmidt, K., {et~al.} 2013, The Astrophysical
  Journal, 775, 98, \dodoi{10.1088/0004-637x/775/2/98}

\bibitem[{{Bruel} {et~al.}(2018){Bruel}, {Burnett}, {Digel}, {Johannesson},
  {Omodei}, \& {Wood}}]{Pass8_Bruel}
{Bruel}, P., {Burnett}, T.~H., {Digel}, S.~W., {et~al.} 2018, arXiv e-prints,
  arXiv:1810.11394, \dodoi{10.48550/arXiv.1810.11394}

\bibitem[{Carilli \& Taylor(2002)}]{Carilli_2002}
Carilli, C.~L., \& Taylor, G.~B. 2002, Annual Review of Astronomy and
  Astrophysics, 40, 319–348, \dodoi{10.1146/annurev.astro.40.060401.093852}

\bibitem[{Carpio {et~al.}(2023)Carpio, Kheirandish, \& Murase}]{Carpio:2022sml}
Carpio, J.~A., Kheirandish, A., \& Murase, K. 2023, JCAP, 04, 019,
  \dodoi{10.1088/1475-7516/2023/04/019}

\bibitem[{Carretti {et~al.}(2022)}]{Carretti:2022tbj}
Carretti, E., {et~al.} 2022, Mon. Not. Roy. Astron. Soc., 512, 945,
  \dodoi{10.1093/mnras/stac384}

\bibitem[{{Charlot} {et~al.}(2020){Charlot}, {Jacobs}, {Gordon}, {Lambert}, {de
  Witt}, {B{\"o}hm}, {Fey}, {Heinkelmann}, {Skurikhina}, {Titov}, {Arias},
  {Bolotin}, {Bourda}, {Ma}, {Malkin}, {Nothnagel}, {Mayer}, {MacMillan},
  {Nilsson}, \& {Gaume}}]{ICRF_radio}
{Charlot}, P., {Jacobs}, C.~S., {Gordon}, D., {et~al.} 2020, \aap, 644, A159,
  \dodoi{10.1051/0004-6361/202038368}

\bibitem[{{Chuprikov} {et~al.}(2004){Chuprikov}, {Ishwara Chandra}, {Guirin},
  \& {Tsarevsky}}]{microquasar_candidate}
{Chuprikov}, A., {Ishwara Chandra}, C.~H., {Guirin}, I., \& {Tsarevsky}, G.
  2004, Astronomical and Astrophysical Transactions, 23, 447,
  \dodoi{10.1080/1055679042000272695}

\bibitem[{{Condon} {et~al.}(1998){Condon}, {Cotton}, {Greisen}, {Yin},
  {Perley}, {Taylor}, \& {Broderick}}]{NVSS}
{Condon}, J.~J., {Cotton}, W.~D., {Greisen}, E.~W., {et~al.} 1998, \aj, 115,
  1693, \dodoi{10.1086/300337}

\bibitem[{Crnogorčević {et~al.}(2025)Crnogorčević, Blanco, \&
  Linden}]{Milena_Cascades}
Crnogorčević, M., Blanco, C., \& Linden, T. 2025, Looking for the
  {$\gamma$}-Ray Cascades of the KM3-230213A Neutrino Source.
\newblock \doarXiv{2503.16606}

\bibitem[{{Dai} {et~al.}(2002){Dai}, {Zhang}, {Gou}, {M{\'e}sz{\'a}ros}, \&
  {Waxman}}]{2002ApJ...580L...7D}
{Dai}, Z.~G., {Zhang}, B., {Gou}, L.~J., {M{\'e}sz{\'a}ros}, P., \& {Waxman},
  E. 2002, \apjl, 580, L7, \dodoi{10.1086/345494}

\bibitem[{{Das} {et~al.}(2025){Das}, {Razzaque}, {Mirabal}, {Omodei}, {Murase},
  \& {Martinez-Castellanos}}]{2025arXiv250415890D}
{Das}, S., {Razzaque}, S., {Mirabal}, N., {et~al.} 2025, arXiv e-prints,
  arXiv:2504.15890, \dodoi{10.48550/arXiv.2504.15890}

\bibitem[{Das {et~al.}(2025)Das, Zhang, Razzaque, \& Xu}]{Das_2025}
Das, S., Zhang, B., Razzaque, S., \& Xu, S. 2025, The Astrophysical Journal,
  991, 96, \dodoi{10.3847/1538-4357/adf8de}

\bibitem[{Dermer {et~al.}(2012)Dermer, Murase, \& Takami}]{Dermer:2012rg}
Dermer, C.~D., Murase, K., \& Takami, H. 2012, Astrophys. J., 755, 147,
  \dodoi{10.1088/0004-637X/755/2/147}

\bibitem[{D’Abrusco {et~al.}(2014)D’Abrusco, Massaro, Paggi, Smith,
  Masetti, Landoni, \& Tosti}]{Wibrals_2014}
D’Abrusco, R., Massaro, F., Paggi, A., {et~al.} 2014, The Astrophysical
  Journal Supplement Series, 215, 14, \dodoi{10.1088/0067-0049/215/1/14}

\bibitem[{D’Abrusco {et~al.}(2019)D’Abrusco, Álvarez Crespo, Massaro,
  Campana, Chavushyan, Landoni, La~Franca, Masetti, Milisavljevic, Paggi,
  Ricci, \& Smith}]{Wibrals_2019}
D’Abrusco, R., Álvarez Crespo, N., Massaro, F., {et~al.} 2019, The
  Astrophysical Journal Supplement Series, 242, 4,
  \dodoi{10.3847/1538-4365/ab16f4}

\bibitem[{Eskenasy {et~al.}(2023)Eskenasy, Kheirandish, \&
  Murase}]{Eskenasy:2022aup}
Eskenasy, R., Kheirandish, A., \& Murase, K. 2023, Phys. Rev. D, 107, 103038,
  \dodoi{10.1103/PhysRevD.107.103038}

\bibitem[{{Fang} {et~al.}(2025){Fang}, {Halzen}, \&
  {Hooper}}]{2025ApJ...982L..16F}
{Fang}, K., {Halzen}, F., \& {Hooper}, D. 2025, \apjl, 982, L16,
  \dodoi{10.3847/2041-8213/adbbec}

\bibitem[{{Huang} \& {HAWC Collaboration}(2025)}]{HAWC_ul}
{Huang}, D., \& {HAWC Collaboration}. 2025, The Astronomer's Telegram, 17069, 1

\bibitem[{Ichiki {et~al.}(2008)Ichiki, Inoue, \& Takahashi}]{Ichiki:2007nd}
Ichiki, K., Inoue, S., \& Takahashi, K. 2008, Astrophys. J., 682, 127,
  \dodoi{10.1086/588275}

\bibitem[{{KM3NeT Collaboration} {et~al.}(2025{\natexlab{a}}){KM3NeT
  Collaboration}, {MessMapp Group}, {Fermi-LAT Collaboration}, {Owens Valley
  Radio Observatory 40-m Telescope Group}, {SVOM Collaboration}, {Baldini},
  {Buchner}, {Erkenov}, {Globus}, {Merloni}, {Paggi}, {Popkov}, {Porquet},
  {Salvato}, {Sotnikova}, \& {Voitsik}}]{2025arXiv250208484K}
{KM3NeT Collaboration}, {MessMapp Group}, {Fermi-LAT Collaboration}, {et~al.}
  2025{\natexlab{a}}, arXiv e-prints, arXiv:2502.08484,
  \dodoi{10.48550/arXiv.2502.08484}

\bibitem[{{KM3NeT Collaboration} {et~al.}(2025{\natexlab{b}}){KM3NeT
  Collaboration}, {Albert}, {Alhebsi}, {Alshamsi}, {Alves Garre}, {Ambrosone},
  {Ameli}, {Andre}, {Anghinolfi}, {Aphecetche}, {Ardid}, {Ardid},
  {Arg{\"u}elles}, {Atmani}, {Aublin}, {Badaracco}, {Bailly-Salins},
  {Barda{\v{c}}ov{\'a}}, {Baret}, {Bariego-Quintana}, {Becherini}, {Bendahman},
  {Benfenati Gualandi}, {Benhassi}, {Bennani}, {Benoit}, {Berbee}, {Bertin},
  {Biagi}, {Boettcher}, {Bonanno}, {Bouasla}, {Boumaaza}, {Bouta}, {Bouwhuis},
  {Bozza}, {Bozza}, {Br{\^a}nza{\c{s}}}, {Bretaudeau}, {Breuhaus}, {Bruijn},
  {Brunner}, {Bruno}, {Buis}, {Buompane}, {Buson}, {Busto}, {Caiffi}, {Calvo},
  {Capone}, {Carenini}, {Carretero}, {Cartraud}, {Castaldi}, {Cecchini},
  {Celli}, {Cerisy}, {Chabab}, {Chen}, {Cherubini}, {Chiarusi}, {Circella},
  {Cocimano}, {Coelho}, {Coleiro}, {Colonges}, {Condorelli}, {Coniglione},
  {Coyle}, {Creusot}, {Cuttone}, {D'Amico}, {Dallier}, {De Benedittis}, {De
  Martino}, {De Wasseige}, {Decoene}, {Del Rosso}, {Di Mauro}, {Di Palma},
  {Diaz}, {Diego-Tortosa}, {Distefano}, {Domi}, {Donzaud}, {Dornic},
  {Drakopoulou}, {Drouhin}, {Ducoin}, {Dvornick{\'y}}, {Eberl}, {Eckerov{\'a}},
  {Eddymaoui}, {van Eeden}, {Eff}, {van Eijk}, {El Bojaddaini}, {El Hedri},
  {Ellajosyula}, {Enzenh{\"o}fer}, {Ferrara}, {Filipovi{\'c}v}, {Filippini},
  {Franciotti}, {Fusco}, {Gagliardini}, {Gal}, {Garc{\'\i}a M{\'e}ndez},
  {Garcia Soto}, {Gatius Oliver}, {Gei{\ss}elbrecht}, {Genton}, {Ghaddari},
  {Gialanella}, {Gibson}, {Giorgio}, {Goos}, {Goswami}, {Gozzini}, {Gracia},
  {Graf}, {Guidi}, {Guillon}, {Guti{\'e}rrez}, {Haack}, {van Haren},
  {Heijboer}, {Hennig}, {Henry}, {Hern{\'a}ndez-Rey}, {Idrissi Ibnsalih},
  {Ilioni}, {Illuminati}, {Joly}, {de Jong}, {de Jong}, {Jung},
  {Kalaczy{\'n}ski}, {Kalekin}, {Kamp}, {Katz}, {Kistauri}, {Kopper},
  {Kouchner}, {Kovalev}, {Kueviakoe}, {Kulikovskiy}, {Kvatadze}, {Labalme},
  {Lahmann}, {Lamoureux}, {Lancelin}, {Larosa}, {Lastoria}, {Lazar}, {Lazo},
  {Le Stum}, {Lehaut}, {Lemaitre}, {Leonora}, {Lessing}, {Levi}, {Lincetto},
  {Lindsey Clark}, {Longhitano}, {Lumb}, {Magnani}, {Majumdar}, {Malerba},
  {Mamedov}, {Manfreda}, {Marconi}, {Margiotta}, {Marinelli}, {Markou},
  {Martin}, {Marzaioli}, {Mastrodicasa}, {Mastroianni}, {Mauro}, {Miele},
  {Migliozzi}, {Migneco}, {Mitsou}, {Mollo}, {Mongelli}, {Morales-Gallegos},
  {Moussa}, {Mozun Mateo}, {Muller}, {Musone}, {Musumeci}, {Navas},
  {Nayerhoda}, {Nicolau}, {Nkosi}, {Fearraigh}, {Oliviero}, {Orlando}, \&
  {Oukacha}}]{2025Natur.638..376K}
{KM3NeT Collaboration}, S., A., {Albert}, A., {Alhebsi}, A.~R., {et~al.}
  2025{\natexlab{b}}, \nat, 638, 376, \dodoi{10.1038/s41586-024-08543-1}

\bibitem[{{Lacy} {et~al.}(2020){Lacy}, {Baum}, {Chandler}, {Chatterjee},
  {Clarke}, {Deustua}, {English}, {Farnes}, {Gaensler}, {Gugliucci},
  {Hallinan}, {Kent}, {Kimball}, {Law}, {Lazio}, {Marvil}, {Mao}, {Medlin},
  {Mooley}, {Murphy}, {Myers}, {Osten}, {Richards}, {Rosolowsky}, {Rudnick},
  {Schinzel}, {Sivakoff}, {Sjouwerman}, {Taylor}, {White}, {Wrobel},
  {Andernach}, {Beasley}, {Berger}, {Bhatnager}, {Birkinshaw}, {Bower},
  {Brandt}, {Brown}, {Burke-Spolaor}, {Butler}, {Comerford}, {Demorest}, {Fu},
  {Giacintucci}, {Golap}, {G{\"u}th}, {Hales}, {Hiriart}, {Hodge}, {Horesh},
  {Ivezi{\'c}}, {Jarvis}, {Kamble}, {Kassim}, {Liu}, {Loinard}, {Lyons},
  {Masters}, {Mezcua}, {Moellenbrock}, {Mroczkowski}, {Nyland}, {O'Dea},
  {O'Sullivan}, {Peters}, {Radford}, {Rao}, {Robnett}, {Salcido}, {Shen},
  {Sobotka}, {Witz}, {Vaccari}, {van Weeren}, {Vargas}, {Williams}, \&
  {Yoon}}]{VLASS}
{Lacy}, M., {Baum}, S.~A., {Chandler}, C.~J., {et~al.} 2020, \pasp, 132,
  035001, \dodoi{10.1088/1538-3873/ab63eb}

\bibitem[{{Mart{\'\i}-Devesa} \& {Olivera-Nieto}(2025)}]{2025ApJ...979L..40M}
{Mart{\'\i}-Devesa}, G., \& {Olivera-Nieto}, L. 2025, \apjl, 979, L40,
  \dodoi{10.3847/2041-8213/ada14f}

\bibitem[{Merloni {et~al.}(2024)Merloni, Lamer, Liu, Ramos-Ceja, Brunner,
  Bulbul, Dennerl, Doroshenko, Freyberg, Friedrich, Gatuzz, Georgakakis,
  Haberl, Igo, Kreykenbohm, Liu, Maitra, Malyali, Mayer, Nandra, Predehl,
  Robrade, Salvato, Sanders, Stewart, Tubín-Arenas, Weber, Wilms, Arcodia,
  Artis, Aschersleben, Avakyan, Aydar, Bahar, Balzer, Becker, Berger, Boller,
  Bornemann, Brüggen, Brusa, Buchner, Burwitz, Camilloni, Clerc, Comparat,
  Coutinho, Czesla, Dannhauer, Dauner, Dauser, Dietl, Dolag, Dwelly, Egg, Ehl,
  Freund, Friedrich, Gaida, Garrel, Ghirardini, Gokus, Grünwald, Grandis,
  Grotova, Gruen, Gueguen, Hämmerich, Hamaus, Hasinger, Haubner, Homan,
  Ider~Chitham, Joseph, Joyce, König, Kaltenbrunner, Khokhriakova, Kink,
  Kirsch, Kluge, Knies, Krippendorf, Krumpe, Kurpas, Li, Liu, Locatelli,
  Lorenz, Müller, Magaudda, Mannes, McCall, Meidinger, Michailidis, Migkas,
  Muñoz-Giraldo, Musiimenta, Nguyen-Dang, Ni, Olechowska, Ota, Pacaud, Pasini,
  Perinati, Pires, Pommranz, Ponti, Poppenhaeger, Pühlhofer, Rau, Reh,
  Reiprich, Roster, Saeedi, Santangelo, Sasaki, Schmitt, Schneider, Schrabback,
  Schuster, Schwope, Seppi, Serim, Shreeram, Sokolova-Lapa, Starck, Stelzer,
  Stierhof, Suleimanov, Tenzer, Traulsen, Trümper, Tsuge, Urrutia, Veronica,
  Waddell, Willer, Wolf, Yeung, Zainab, Zangrandi, Zhang, Zhang, \&
  Zheng}]{eROSITA_DR1}
Merloni, A., Lamer, G., Liu, T., {et~al.} 2024, Astronomy amp; Astrophysics,
  682, A34, \dodoi{10.1051/0004-6361/202347165}

\bibitem[{{Mirabal}(2023)}]{2023MNRAS.519L..85M}
{Mirabal}, N. 2023, Mon. Not. R. Astron. Soc., 519, L85,
  \dodoi{10.1093/mnrasl/slac157}

\bibitem[{Moderksi {et~al.}(2005)Moderksi, Sikora, Coppi, \&
  Aharonian}]{Moderksi:2005jw}
Moderksi, R., Sikora, M., Coppi, P.~S., \& Aharonian, F.~A. 2005, Mon. Not.
  Roy. Astron. Soc., 363, 954, \dodoi{10.1111/j.1365-2966.2005.09814.x}

\bibitem[{Murase(2009)}]{Murase:2009ah}
Murase, K. 2009, Phys. Rev. Lett., 103, 081102,
  \dodoi{10.1103/PhysRevLett.103.081102}

\bibitem[{Murase(2012)}]{Murase:2011yw}
---. 2012, Astrophys. J. Lett., 745, L16, \dodoi{10.1088/2041-8205/745/2/L16}

\bibitem[{Murase {et~al.}(2008)Murase, Takahashi, Inoue, Ichiki, \&
  Nagataki}]{Murase:2008pe}
Murase, K., Takahashi, K., Inoue, S., Ichiki, K., \& Nagataki, S. 2008,
  Astrophys. J. Lett., 686, L67, \dodoi{10.1086/592997}

\bibitem[{{Plaga}(1995)}]{1995Natur.374..430P}
{Plaga}, R. 1995, \nat, 374, 430, \dodoi{10.1038/374430a0}

\bibitem[{Prokhorov \& Moraghan(2022)}]{Prokhorov_2022}
Prokhorov, D.~A., \& Moraghan, A. 2022, Monthly Notices of the Royal
  Astronomical Society, 519, 2680–2689, \dodoi{10.1093/mnras/stac3453}

\bibitem[{{Skrutskie} {et~al.}(2006){Skrutskie}, {Cutri}, {Stiening},
  {Weinberg}, {Schneider}, {Carpenter}, {Beichman}, {Capps}, {Chester},
  {Elias}, {Huchra}, {Liebert}, {Lonsdale}, {Monet}, {Price}, {Seitzer},
  {Jarrett}, {Kirkpatrick}, {Gizis}, {Howard}, {Evans}, {Fowler}, {Fullmer},
  {Hurt}, {Light}, {Kopan}, {Marsh}, {McCallon}, {Tam}, {Van Dyk}, \&
  {Wheelock}}]{2MASS}
{Skrutskie}, M.~F., {Cutri}, R.~M., {Stiening}, R., {et~al.} 2006, \aj, 131,
  1163, \dodoi{10.1086/498708}

\bibitem[{Takahashi {et~al.}(2008)Takahashi, Murase, Ichiki, Inoue, \&
  Nagataki}]{Takahashi:2008pc}
Takahashi, K., Murase, K., Ichiki, K., Inoue, S., \& Nagataki, S. 2008,
  Astrophys. J. Lett., 687, L5, \dodoi{10.1086/593118}

\bibitem[{{Waxman} \& {Coppi}(1996)}]{1996ApJ...464L..75W}
{Waxman}, E., \& {Coppi}, P. 1996, \apjl, 464, L75, \dodoi{10.1086/310090}

\bibitem[{{Wenger} {et~al.}(2000){Wenger}, {Ochsenbein}, {Egret}, {Dubois},
  {Bonnarel}, {Borde}, {Genova}, {Jasniewicz}, {Lalo{\"e}}, {Lesteven}, \&
  {Monier}}]{Simbad}
{Wenger}, M., {Ochsenbein}, F., {Egret}, D., {et~al.} 2000, \aaps, 143, 9,
  \dodoi{10.1051/aas:2000332}

\bibitem[{Wood {et~al.}(2017)Wood, Caputo, Charles, Mauro, Magill, \&
  Perkins}]{fermipy}
Wood, M., Caputo, R., Charles, E., {et~al.} 2017, Fermipy: An open-source
  Python package for analysis of Fermi-LAT Data.
\newblock \doarXiv{1707.09551}

\bibitem[{{Wright} {et~al.}(2010){Wright}, {Eisenhardt}, {Mainzer}, {Ressler},
  {Cutri}, {Jarrett}, {Kirkpatrick}, {Padgett}, {McMillan}, {Skrutskie},
  {Stanford}, {Cohen}, {Walker}, {Mather}, {Leisawitz}, {Gautier}, {McLean},
  {Benford}, {Lonsdale}, {Blain}, {Mendez}, {Irace}, {Duval}, {Liu}, {Royer},
  {Heinrichsen}, {Howard}, {Shannon}, {Kendall}, {Walsh}, {Larsen}, {Cardon},
  {Schick}, {Schwalm}, {Abid}, {Fabinsky}, {Naes}, \& {Tsai}}]{WISE}
{Wright}, E.~L., {Eisenhardt}, P. R.~M., {Mainzer}, A.~K., {et~al.} 2010, \aj,
  140, 1868, \dodoi{10.1088/0004-6256/140/6/1868}

\end{thebibliography}

\appendix

\section{Spectral energy distribution of sub-threshold sources} \label{appendix:spectra}

Here we present the spectral energy distributions of the three sub-threshold Fermi-LAT sources in the vicinity of KM3-230213A. The spectra are evaluated on 500 MeV - 1 TeV. The spectra are evaluated using the full $\sim 17$ year dataset for J0614.6-0731 and J0621.1-0610, while the spectrum for J0616.1-0428 is evaluated on the $\sim 2$ year dataset following KM3-230213A. The initial fit is performed after adding the sub-threshold sources to the model and freeing all parameters. The spectral energy distributions are evaluated using \texttt{fermipy} SED fitting function \texttt{gta.sed}, which assumes a powerlaw spectrum with spectral index $\Gamma = 2$ within each energy bin. In the legend we also list the overall best-fit spectral index.

\begin{figure}[h]
     \centering
    \includegraphics[width=0.44\textwidth]{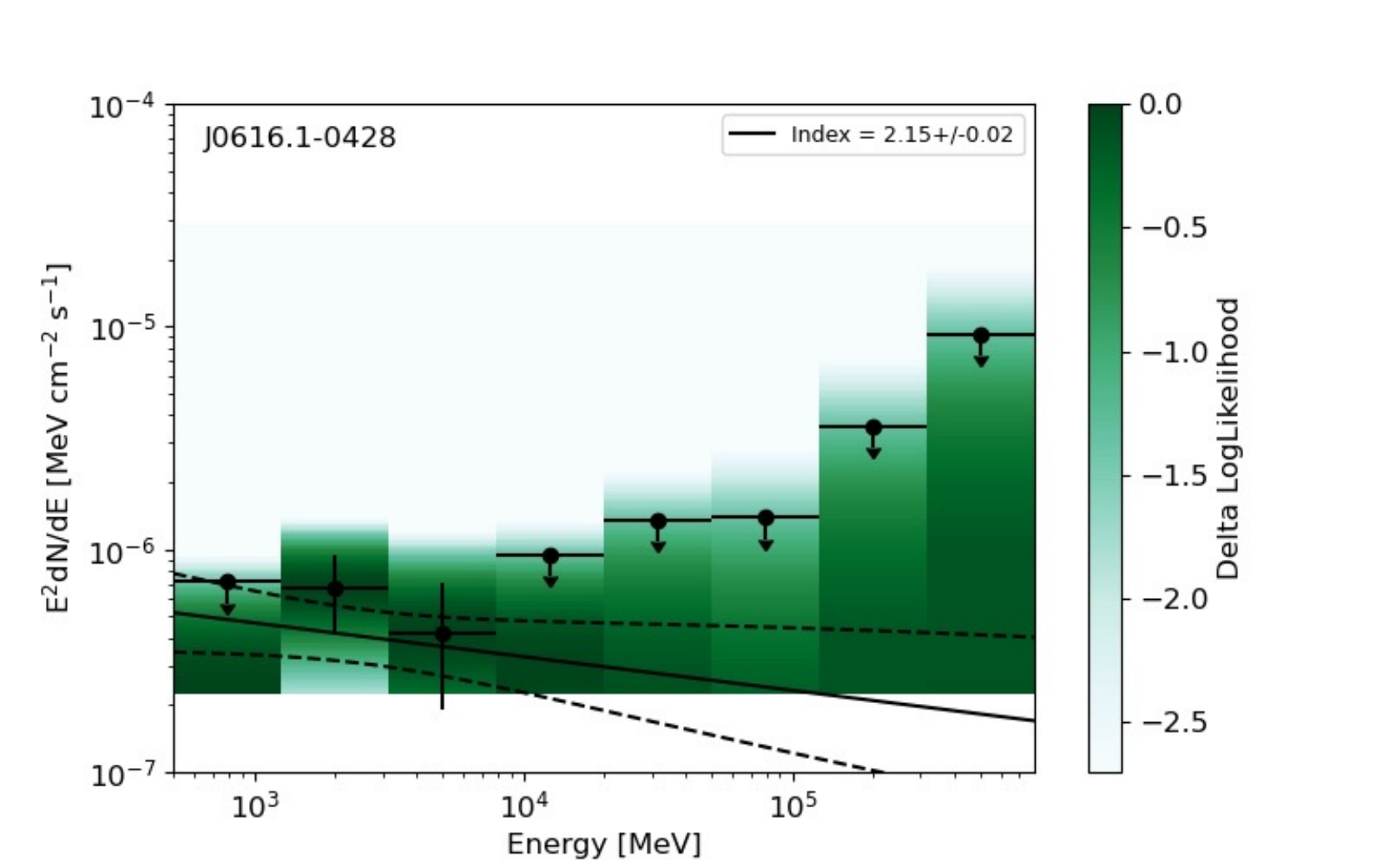}
    \includegraphics[width=0.44\textwidth]{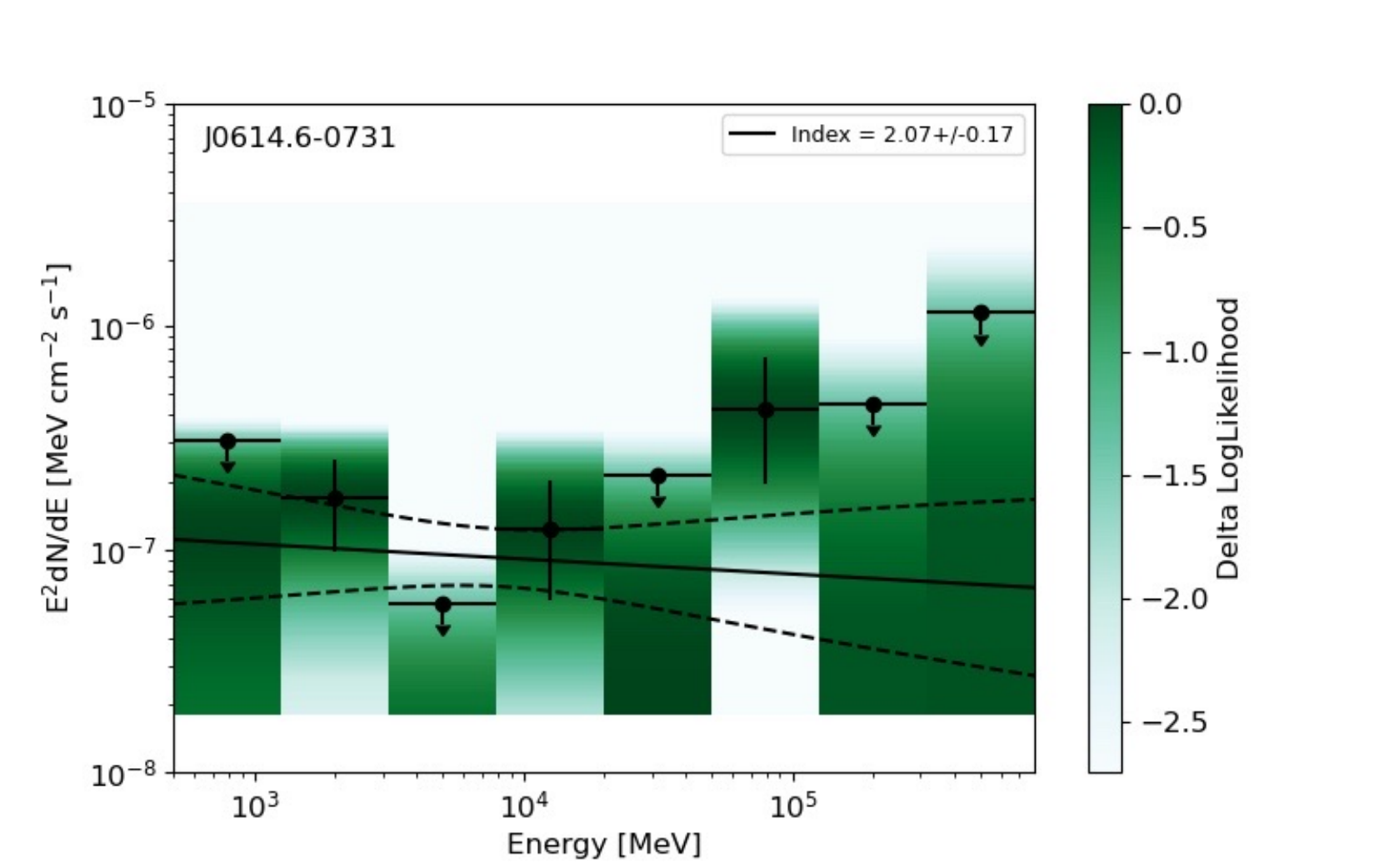}
    \includegraphics[width=0.44\textwidth]{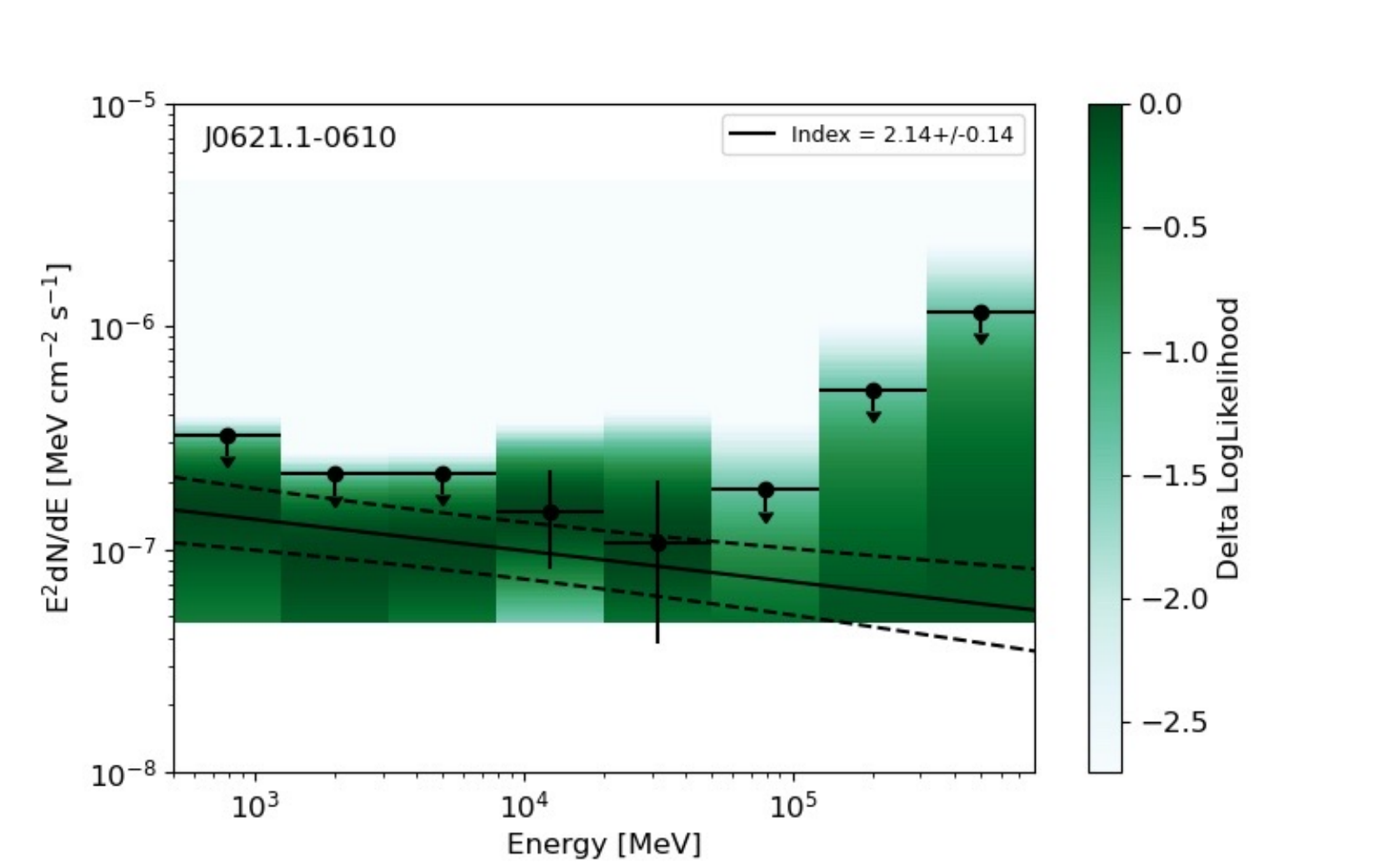}]
    \caption{Spectral energy distributions (SEDs) for the three sub-threshold sources uncovered in this analysis.}
    \label{spectra_alltime}
\end{figure}

\end{document}